\documentclass[%
 reprint,
 amsmath,amssymb,
 aps,
]{revtex4-1}
\usepackage{graphicx}% Include figure files
\usepackage{epstopdf}
\usepackage{dcolumn}% Align table columns on decimal point
\usepackage{bm}% bold math
\usepackage{balance}
\usepackage{multirow}
\begin{document}

%\preprint{APS/123-QED}

\title{Resonant high-energy bremsstrahlung of ultrarelativistic electrons in the field of a nucleus and a pulsed light wave}% Force line breaks with \\
%%\thanks{A footnote to the article title}%

\author{Sergei P. Roshchupkin}
 \email{serg9rsp@gmail.com}
\affiliation{
 Department of Theoretical Physics, Peter the Great St. Petersburg Polytechnic University\\
 St-Petersburg, Russia}

\author{Alexander Dubov}
 \email{alexanderpolytech@gmail.com}
\affiliation{
 Department of Theoretical Physics, Peter the Great St. Petersburg Polytechnic University\\
St-Petersburg, Russia}
\affiliation{
Department of Applied Physics, Aalto University\\
Helsinki, Finland}

\author{Victor V. Dubov}
 \email{dubov@spbstu.ru}
 \affiliation{
 Department of Theoretical Physics, Peter the Great St. Petersburg Polytechnic University\\
 St-Petersburg, Russia}

\date{\today}% It is always \today, today,

\begin{abstract}
The actual theoretical research investigates the resonant high-energy spontaneous bremsstrahlung of ultrarelativistic electrons with considerable energies $E_i \gtrsim 10^2 GeV$ in the field of a nucleus and a quasimonochromatic laser wave with intensity $I \lesssim 10^{16} \div 10^{17} W/cm^2$. Under the resonant conditions within the laser field the intermediate virtual electron transforms into the real particle. Therefore, the initial second-order process with accordance to the fine structure constant effectively splits into two consequent first-order phenomena: the laser-stimulated Compton-effect and laser field-assisted scattering of an electron on a nucleus. As a result, the accomplished analysis defines that the polar emission angle characterizes the frequency of a spontaneous photon. The study derives the expressions for the resonant differential cross-sections of the represented processes that realize simultaneous registration of the frequency and radiation angle in correlation to the momentum of the initial electron (for the channel A) and of the final electron (for the channel B) of the spontaneous photon with absorption of $r$ wave photons ($r = 1, 2, 3,... $ - the number of a resonance). Additionally, the distribution of the resonant differential cross-section as a function of the angle of the spontaneous photon emission for the higher numbers of resonance ($r = 2, 3,... $) delineates a dependency with a sharp peak maximum that coordinates to the particle radiation at the most probable frequency. To summarize, the accomplished work represents that the resonant differential cross-section acquires considerable magnitude. Thus, for the first resonance of the channel A the resonant differential cross-section attains the $\sim 10^{12}$ order of a magnitude, and for the third resonance of the channel B $\sim 10^5$ order of a magnitude (in the units of $\alpha Z^2 r_e^2$). Finally, numerous scientific facilities with specialization in pulsed laser radiation (SLAC, FAIR, XFEL, ELI, XCELS) may experimentally verify the constructed model calculations.
\end{abstract}

%Use showkeys class option if keyword
\maketitle

%\tableofcontents

\section{\label{sec:level1}Introduction}
The advancement of the contemporary industry of powerful sources of laser radiation establishes an essential demand for the fundamental research programs \cite{1,2,3,4,5}. Consequently, the theoretical study of the processes of quantum electrodynamics (QED) within the strong light fields represents an intensively developing area of investigations that accumulates a significant priority level (see, for example, articles \cite{6,7,8,9,10,11,12,13,14,15,16,17,18,19,20,21,22,23,24,25,26,27,28,29,30,31,32,33,34,35,36,37,38,39,40,41,42,43,44,45,46,47,48,49}). Therefore, monographs \cite{11,12,13,14} and reviews \cite{15,16,17,18,19,20,21} systematize principal results. In addition, it is important to underline that the high-order QED processes with accordance to the fine structure constant in the light field (laser-assisted QED processes) occur within the non-resonant and resonant channels. The Oleinik resonances \cite{9,10} that appear in the laser field correspond to the proposition that the lower-order QED effects with accordance to the fine structure constant (laser-stimulated QED processes) \cite{15} can materialize in the light field ambience. Consequently, the probability of the resonant development of the QED processes substantially exceeds (by several orders of magnitude) the coordinate probability for the phenomenon proceeding in the absence of the external field \cite{19,20}.\\
The works \cite{12,13,14,19,20,47} scrutinize the Oleinik resonances for the case of spontaneous bremsstrahlung (SB) of an electron on a nucleus in the field of an electromagnetic wave. However, the articles \cite{12,13,14,19,20} analyze the resonances for a single scheme of reaction distinctively - when the electron radiates a spontaneous photon and then scatters on a nucleus. The second interaction - when an electron scatters on a nucleus and subsequently emits a spontaneous photon constitutes a proposition for the actual investigation. Furthermore, the examination of the first reaction channel in the ultrarelativistic case implemented only the moderate energies of electrons that scattered at considerable angles and the spontaneous photon emission in a narrow cone along the direction of the initial electron. The paper \cite{47} evaluates the resonances for two reaction schemes for the model with considerable energies of ultrarelativistic electrons in the field of a plane monochromatic electromagnetic wave. The article \cite{48} represents the cross-channel of the indicated reactions - the resonant photoproduction of ultrarelativistic pairs by a hard gamma-quantum in the field of a nucleus and a plane monochromatic wave. The articles \cite{47,48} implement the Breit-Wigner procedure in order to eliminate the resonant infinity with evaluation of the corresponding radiation width.\\
The actual paper focuses on the theory of high-energy bremsstrahlung of ultrarelativistic electrons with considerable energies $E_i \gtrsim 10^2 GeV$ under the condition of scattering in the Coulomb field of a nucleus in the presence of an external pulsed laser field with intensity $I \lesssim 10^{16} \div 10^{17} W/cm^2$. Additionally, it is important to underline that in the pulsed laser field the resonant width appears as the consequence of the mathematical formalism \cite{20}.\\
The scrutiny of the SB phenomenon delineates a couplet of characteristic parameters. The classical relativisticaly-invariant parameter \cite{15,19,20,21,22}:
\begin{equation} \label{eq:1}
\eta_0  = \frac{{eF_0\mathchar'26\mkern-10mu\lambda  }}{{m{c^2}}}
\end{equation}
that is equal to the ratio of the work of the field at a wavelength to the rest energy of an electron (where $e$ and $m$ - are the charge and mass of an electron, $F_0$ and $\mathchar'26\mkern-10mu\lambda = c/\omega$ - are the electric field strenght and the wavelength, and $\omega$ - is the frequency of a wave). Consequently, the quantum parameter (Bunkin-Fedorov parameter) \cite{8,11,19,20,21,22}:
\begin{equation} \label{eq:2}
\gamma_0 = \eta_0 \frac {m v_i c}{\hbar \omega}
\end{equation}
Within the range of the optical frequencies ($\omega \sim 10^{15} s^{-1}$) the classical parameter is $\eta_0 \sim 1$ for the fields $F_0 \sim 10^{10} \div 10^{11} V/cm$ and the quantum parameter is $\gamma_0 \sim 1$ for the fields $F_0 \sim (10^5 \div 10^6)(c/v_i) V/cm$. Therefore,  in the represented ambience the $\gamma_0 \gg \eta_0$. Thus, the general parameter that defines multiphoton processes in the SB procedure of electron scattering at considerable angles is the quantum Bunkin-Fedorov parameter. However, under the condition of scattering of electrons on a nucleus at small-scaled angles the quantum parameter $\gamma_0$ \eqref{eq:2} does not manifest significantly and the primal parameter that research utilizes for the evaluation of the multiphoton effects is the classical parameter $\eta_0$ \eqref{eq:1} \cite{44,47}.\\
The application of the powerful ultrashort laser pulses in the contemporary experiments proposes a specific interest for examination of the various effects emanating within the represented light fields. The amplitude of the field strength of the pulsed laser varies significantly both in space and in time scales. The standard approach for the construction of an appropriate model of interactions of electrons with the pulsed laser field utilizes the plane non-monochromatic wave configuration with the characteristic pulse duration that coordinates with the condition \cite{7,20}:
\begin{equation} \label{eq:3}
\omega \tau \gg 1
\end{equation}
In the range of the optical frequencies the characteristic pulse duration may constitute tens of femtoseconds and, subsequently, the expression \eqref{eq:3} is appropriate for the modern high-power pulsed laser sources. The fields that correlate to the condition \eqref{eq:3} are referred as quasimonochromatic. The actual study considers the ultrarelativistic electrons and weak quasimonochromatic fields \eqref{eq:3} - the indicated conditions delineate that the electron does not deviate from the initial direction of propagation as a result of the pomderomotive effect in the light field. For the represented arrangement the investigation neglects the spatial inhomogeneity of the beam in the transverse direction from the examination \cite{7,20}.\\
The accomplished analysis in the article utilizes relativistic system of units: $\hbar=c=1$.

\section{\label{sec:level1}The amplitude of the process}

\begin{figure}[h!]
     \begin{center}
     \includegraphics[width=8cm]{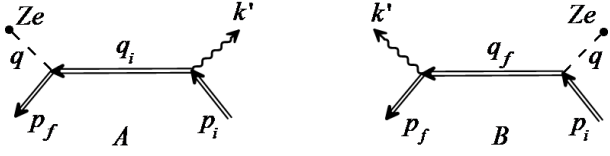}
     \end{center}
     \caption{Feynman diagrams of the process of SB of an electron on a nucleus in the field of a pulsed light wave. The double incoming and outgoing lines delineate the Wolkow wave functions of an electron in the initial and final states, the inner double line designates the Green function of an electron within a pulsed light field \eqref{eq:4}. The waved lines  coordinate to the 4-momenta of the spontaneous photon and the dashed lines illustrate the "pseudo-photon" of the nucleus recoil.}
     \label{Fig:Figure 1}
  \end{figure}

The research selects the following form of the 4-potential of the plane-wave electromagnetic impulse that propagates along the $z$ axis:
\begin{equation} \label{eq:4}
\begin{aligned}
A({\phi}) = A_0 {\cdot} g({\frac {\varphi} {\omega \tau}}) {\cdot} (e_x \cos{\varphi} + \delta e_y \sin{\varphi}), \\ \varphi=kx={\omega(t-z)}
\end{aligned}
\end{equation}
Here $A_0 = F_0/\omega$, $k=(\omega, {\bf k})$ - is the 4-vector of a wave; $F_0$, $\omega$ and $\delta$ - are the field strength, frequency and the ellipticity parameter of polarization of a wave; $e_{x}=(0,{\bf e}_{x}), \: e_{y}=(0,{\bf e}_{y})$ - are the 4-vectors of wave polarization, where $e_{x,y}^2=-1, \: (e_{x,y}k)=k^2=0$. In the expression \eqref{eq:4} the function $g(\varphi/\omega \tau)$ - is the envelope function of the potential and it is equal to $g(0) = 1$ at the center of the impulse, and with $g(0) \rightarrow 0$ the envelope function exponentially decreases with the condition: $|\varphi| \gg \omega \tau$. The represented limitations appropriate that the magnitude $\tau$ designates the duration of the laser pulse.\\
The consequential investigation asserts that the pulse duration is essentially longer than the characteristic wave oscillation time \eqref{eq:3}. The examination analyzes the SB of electrons on a nucleus in the pulsed light field in the Born approximation of the electrons with nucleus field interaction ($Ze^2/v \ll 1$, $v$ - is the velocity of an electron, $c$ - is the velocity of light in vacuum, $Z$ - the charge of a nucleus). Two Feynman diagrams illustrate the indicated second with accordance to the fine structure order process (see Fig. \ref{Fig:Figure 1}).

The research allocates the following form for the amplitude of the process of spontaneous bremsstrahlung of an electron on a nucleus in the field of a pulsed light wave:
\begin{equation} \label{eq:5}
\begin{aligned}
S_{fi} = -ie^2 \int d^4 x_1 d^4 x_2 \overline{\psi}_f (x_2 | A) {\cdot} \\ {\cdot} [\tilde{\gamma}_0 A_0 (|{\bf x_2}|)G(x_2 x_1 | A) \hat{A'} (x_1, k') + \\ + \hat{A'} (x_2, k') G(x_2 x_1 | A) \tilde{\gamma}_0 A_0 (|{\bf x_1}|)] \psi_i (x_1 | A)
\end{aligned}
\end{equation}
Where $\psi_i (x_1 | A)$ and $\overline{\psi}_f (x_2 | A)$ - are the wave functions of electron in the initial and final states, and $G(x_2 x_1 | A)$ - is the Green function of an intermediate electron within the field of a pulsed light wave \eqref{eq:4}. Hereinafter, the magnitudes with a hat imply the scalar product expression of the corresponding 4-vector and Dirac gamma-matrix: $\tilde{\gamma}_{\mu} = (\tilde{\gamma}_0, {\bf \tilde{\gamma}})$, $\mu = 0, 1, 2, 3$. For example: $\hat{A'} = A'_{\mu} \tilde{\gamma}^{\mu} = A'_0 \tilde{\gamma}^0 - {\bf A' \tilde{\gamma}}$. In the equation \eqref{eq:5} $A_0 (|{\bf x}_n|)$ - is the Coulomb potential of a nucleus and $A'_{\mu} (x_n, k')$ - is the 4-potential of a spontaneous photon that have the following outline:
\begin{equation} \label{eq:6}
A_0 (|{\bf x}_n|) = {\frac {Z e} {|{\bf x}_n|}}
\end{equation}
\begin{equation} \label{eq:7}
A'_{\mu} (x_n, k') = {\sqrt{\frac {2 \pi} {\omega'}}} \varepsilon^{*}_{\mu} e^{i k' x_n}, \: n = 1, 2
\end{equation}
Where $\varepsilon^{*}_{\mu}$ and $k' = (\omega', {\bf k'})$ - are the 4-vector of polarization and the 4-momentum of a spontaneous photon, $(k' x_n) = (\omega' t_n - {\bf k'} {\bf x}_n)$.
The study determines the envelope function of the potential as being dependant on the variable $\phi$. Therefore, the electromagnetic field \eqref{eq:4} delineates as the plane wave. The Wolkow functions \cite{49,50} are the generally recognized exact solutions of the Dirac equations for the electron in the field of a plane wave of arbitrary spectral composition. Additionally, the research utilizes the model of the Green function \cite{51,52} for the electron within the plane wave. Thus, the scrutiny represents wave functions and the Green function from the statement \eqref{eq:5} in the following form:
\begin{equation} \label{eq:8}
\begin{cases} \psi_i (x_1 | A) = D_i (x_1) \cdot {\frac {u_i} {\sqrt{2 E_i}}}, \\ \overline{\psi}_f (x_2 | A) = {\frac {\overline{u}_f} {\sqrt{2 E_f}}} \cdot \overline{D}_f (x_2); \end{cases}
\end{equation}
\begin{equation} \label{eq:9}
G(x_2 x_1 | A) = \int {\frac {d^4 p} {(2 \pi)^4}} \cdot D_p (x_1) {\frac {\hat{p} + m} {p^2 - m^2}} \overline{D}_p (x_2)
\end{equation}
Where
\begin{equation} \label{eq:10}
\begin{aligned}
D_\varsigma (x_n) \equiv D_{n \varsigma} = [1 + {\frac {e} {2 (k p_\varsigma)}} \hat{k} \cdot \hat{A} (k x_n)] e^{i S (p_\varsigma , x_n)}, \\ n = 1, 2; \: p_\varsigma = p, p_i, p_f
\end{aligned}
\end{equation}
\begin{equation} \label{eq:11}
\begin{aligned}
S (p_\varsigma, x_n) = - (p_\varsigma , x_n) - {\frac {e} {(k p_\varsigma)}} \cdot \\ \cdot \int^{k x_n} [(p_\varsigma A(\varphi')) - {\frac {e} {2}} A^2 (\varphi')] d\varphi'
\end{aligned}
\end{equation}
$p_{i, f} = (E_{i, f}, {\bf p}_{i, f})$ - are the 4-momenta of the initial and final electrons. The equation \eqref{eq:11} is the classical action of an electron in the external light field.
The sequential investigation considers the examined process for the arrangement with circular polarization $(\delta^2 = 1)$ and the field intensity value approximated at the peak of the pulse
\begin{equation} \label{eq:12}
\eta_0 \ll 1
\end{equation}
With application of various analytical and computational methods (see, for example \cite{20}) the paper derives the particular construct for the amplitude of the SB of electrons on a nucleus within the field of a weak plane quasimonochromatic \eqref{eq:3} electromagnetic wave \eqref{eq:13}:
\begin{equation} \label{eq:13}
S_{fi} = \sum_{l=-\infty}^{\infty} S_{(l)}
\end{equation}
Where $S_{(l)}$ - is the partial amplitude of the described effect with the emission (absorption) of $|l|$ wave photons.
\begin{equation} \label{eq:14}
S_{(l)} = -i {\frac {2 \tau^2 \omega Z e^3 \sqrt{\pi}} {\sqrt{2 \omega' E_f E_i}}} {\frac {(\overline{u}_f B_{(l)} u_i)} {[{\bf q}^2 + q_0 (q_0 - 2 q_z)]}}
\end{equation}
Here
\begin{equation} \label{eq:15}
B_{(l)} = B_{i(l)} + B_{f(l)}
\end{equation}
\begin{equation} \label{eq:16}
q = p_f - p_i + k' +lk
\end{equation}
Where the $q = (q_0, {\bf q})$ - is the transmitted 4-momentum, $B_{i(l)}$ and $B_{f(l)}$ are the amplitudes for the channels A and B.
\begin{equation} \label{eq:17}
\begin{aligned}
B_{i(l)} = \sum_{r=-\infty}^{\infty} \int_{-\infty}^{\infty} d \phi_1 \int_{-\infty}^{\infty} d \phi_2 \: e^{(i q_0 \tau \phi_2)} \cdot \\ \cdot M_{(l+r)}^0 (p_f, q_i, \phi_2) G (q_i, \phi_1 - \phi_2) [\varepsilon_{\mu}^{*} F_{(-r)}^{\mu} (q_i, p_i, \phi_1)]
\end{aligned}
\end{equation}
\begin{equation} \label{eq:18}
\begin{aligned}
B_{f(l)} = \sum_{r=-\infty}^{\infty} \int_{-\infty}^{\infty} d \phi_1 \int_{-\infty}^{\infty} d \phi_2 \: e^{(i q_0 \tau \phi_1)} \cdot \\ \cdot [\varepsilon_{\mu}^{*} F_{(-r)}^{\mu} (p_f, q_f, \phi_2)] G (q_f, \phi_2 - \phi_1) M_{(l+r)}^0 (q_f, p_i, \phi_1)
\end{aligned}
\end{equation}
\begin{equation} \label{eq:19}
\begin{aligned}
G (q_i, \phi_1 - \phi_2) = \int_{-\infty}^{\infty} d \xi [{\frac {(\hat{q}_i + m) + \xi \hat{k}} {(q_i^2-m^2) + 2 \xi (k q_i)}}] \cdot \\ \cdot e^{i (\omega \tau \xi) (\phi_1 - \phi_2)}
\end{aligned}
\end{equation}
\begin{equation} \label{eq:20}
q_i = p_i - k' + r k, q_f = p_f + k' - r k
\end{equation}
The addend $G (q_i, \phi_2 - \phi_1)$ in the expression \eqref{eq:18} derives from the ratio \eqref{eq:19} with substitution: $q_i \rightarrow q_f$, $\phi_1 \leftrightarrow \phi_2$. In addition, $\phi_n = \varphi_n / \omega \tau (n = 1, 2)$, $q_i$ and $q_f$ - are the 4-momenta of the intermediate electrons for channels A and B (see Fig. \ref{Fig:Figure 1} and Fig. \ref{Fig:Figure 2}). In the equations \eqref{eq:17} and \eqref{eq:18} under the resonant conditions and with accordance to the sum in the denominator within the square brackets of \eqref{eq:14}: $M_{(l + r)}^0 (p', p, \phi_n)$ - the amplitude of the electron scattering on a nucleus $(p \rightarrow p')$ with radiation (absorption) of $|l + r|$ - photons of the wave (laser field-assisted Mott process) \cite{11} and $F_{(-r)}^{\mu} (p', p, \phi_n)$ - the amplitude of emission of spontaneous photon $k'$ by an electron $(p \rightarrow p')$ as a result of the absorption of $r$ - photons of a wave (laser field-stimulated Compton effect) \cite{15} obtain a form:
\begin{equation} \label{eq:21}
M_{(l + r)}^0 (p', p, \phi_n) = \tilde{\gamma}^0 L_{(l + r)} (p', p, \phi_n),
\end{equation}
\begin{equation} \label{eq:22}
\begin{aligned}
F_{(-r)}^{\mu} (p', p, \phi_n) = \tilde{\gamma}^{\mu} L_{(-r)} (p', p, \phi_n) + \\ + b_{p'p(-)}^{\mu} (\phi_n) L_{(-r-1)} + b_{p'p(+)}^{\mu} (\phi_n) L_{(-r+1)}, \: n = 1, 2
\end{aligned}
\end{equation}
The matrices $b_{p'p(\pm)}^{\mu}$ from the statements \eqref{eq:21}, \eqref{eq:22} and the special functions $L_{n'} (p', p, \phi_n)$ that the article \cite{46} scrutinized have the structure of:
\begin{equation} \label{eq:23}
b_{p'p(\pm)}^{\mu} (\phi_n) = \eta (\phi_n) [{\frac {m} {4 (k p')}} \hat{e}_{\pm} \hat{k} \tilde{\gamma}^{\mu} + {\frac {m} {4 (k p)}} \tilde{\gamma}^{\mu} \hat{k} \hat{e}_{\pm}]
\end{equation}
\begin{equation} \label{eq:24}
L_s \equiv L_s (p', p, \phi_n) = e^{-i s \chi_{p' p}} J_s [\gamma_{p' p} (\phi_n)]
\end{equation}
Here $J_s$ - are the Bessel functions of an integer index and the parameters $\gamma_{p' p}, \chi_{p' p}$, and 4-vectors $e_{\pm}$ are equal to:
\begin{equation} \label{eq:25}
\gamma_{p' p} (\phi_n) = \eta (\phi_n) m  \sqrt{-Q_{p'p}^2}, \: Q_{p'p} = {\frac {p'} {(k p')}} - {\frac {p} {(k p)}}
\end{equation}
\begin{equation} \label{eq:26}
\tan \chi_{p'p} = \delta \cdot \tan {\frac {(Q_{p'p} e_y)} {(Q_{p'p} e_x)}}, \: e_{\pm} = e_x \pm i \delta e_y
\end{equation}
The expressions for the amplitudes \eqref{eq:17} and \eqref{eq:18} define the values of the 4-momenta $p$ and $p'$ from the equations \eqref{eq:21}-\eqref{eq:26}. The ratio \eqref{eq:19} indicates that the following range designates the substantial area of the integration variable $\xi$:
\begin{equation} \label{eq:27}
|\xi| \lesssim {\frac {1} {\omega \tau}} \ll 1
\end{equation}
The magnitudes of the integrals are moderate as the consequence of the extensive oscillations of the integrand when $|\xi| \gg {\frac {1} {\omega \tau}}$. Therefore, in the numerator of the integrand in the equation \eqref{eq:19} the research neglects $\xi \hat{k}$ from the consideration in contrast to $(\hat{q}_{i, f} + m)$. Moreover, the dependency on the integration variable in the denominator of the expression \eqref{eq:19} is a result of accounting of the pulsed character of the external laser wave \cite{20}. For the model with the monochromatic wave the similar correction is absent and, subsequently, the specified arrangement delineates the resonant divergence of the amplitude of the process of SB of an electron on a nucleus within the field of an electromagnetic wave \cite{19,47}. Thus, the investigation calculates the integral \eqref{eq:19} and formulates patterns for the channels A and B:
\begin{equation} \label{eq:28}
G (q_i, \phi_1 - \phi_2) = {\frac {\pi i (\hat{q}_{i} + m)} {(k q_i)}} e^{-2 i \beta_i (\phi_1 - \phi_2)} {sgn} (\phi_1 - \phi_2)
\end{equation}
\begin{equation} \label{eq:29}
G (q_f, \phi_2 - \phi_1) = {\frac {\pi i (\hat{q}_{f} + m)} {(k q_f)}} e^{-2 i \beta_f (\phi_2 - \phi_1)} {sgn} (\phi_2 - \phi_1)
\end{equation}
Where $\beta_i$ and $\beta_f$ - are the resonant parameters for the channels A and B \cite{20}.
\begin{equation} \label{eq:30}
\beta_i = {\frac {q_i^2 - m^2} {4(k q_i)}} \omega \tau, \: \beta_f = {\frac {q_f^2 - m^2} {4(k q_f)}} \omega \tau,
\end{equation}
With reexamination of \eqref{eq:28}, \eqref{eq:29} the amplitudes \eqref{eq:17} and \eqref{eq:18} obtain a form:
\begin{equation} \label{eq:31}
\begin{aligned}
B_{i(l)} = \sum_{r=-\infty}^{\infty} {\frac {\pi i} {(k q_i)}} \int_{-\infty}^{\infty} d \phi_1 \int_{-\infty}^{\infty} d \phi_2 \:  e^{(-2 i \beta_i \phi_1)} \cdot \\ \cdot e^{i (q_0 \tau + 2 \beta_i) \phi_2} {sgn} (\phi_1 - \phi_2) [M_{(l+r)}^0 (p_f, q_i, \phi_2)] \cdot \\ \cdot (\hat{q}_i + m) [\varepsilon_{\mu}^{*} F_{(-r)}^{\mu} (q_i, p_i, \phi_1)]
\end{aligned}
\end{equation}
\begin{equation} \label{eq:32}
\begin{aligned}
B_{f(l)} = \sum_{r=-\infty}^{\infty} {\frac {\pi i} {(k q_f)}} \int_{-\infty}^{\infty} d \phi_1 \int_{-\infty}^{\infty} d \phi_2 \: e^{(-2 i \beta_f \phi_2)} \cdot \\ \cdot e^{i (q_0 \tau + 2 \beta_f) \phi_1} {sgn} (\phi_2 - \phi_1) [\varepsilon_{\mu}^{*} F_{(-r)}^{\mu} (p_f, q_f, \phi_2)] \cdot \\ \cdot (\hat{q}_f + m) [M_{(l+r)}^0 (q_f, p_i, \phi_1)]
\end{aligned}
\end{equation}
The derived amplitude \eqref{eq:13}-\eqref{eq:15}, \eqref{eq:31}, \eqref{eq:32} is valid for the circular polarization of the weak quasimonochromatic laser wave \eqref{eq:3}, \eqref{eq:12}.\\
The consequential research concentrates on the outlined process within the range of ultrarelativistic energies of electrons and hard spontaneous photons for the configuration of the particles propagation within the narrow angle cone in the direction complementary to the direction of the momentum of the initial electron \cite{47}.
\begin{equation} \label{eq:33}
E_{i, f} \gg m
\end{equation}
\begin{equation} \label{eq:34}
\theta'_{i,f} = \measuredangle ({\bf k}',{\bf p}_{i,f}) \ll 1, \: \overline{\theta}_{if} = \measuredangle ({\bf p}_i,{\bf p}_f) \ll 1
\end{equation}
\begin{equation} \label{eq:35}
\theta' = \measuredangle ({\bf k}',{\bf k}) \sim 1, \: \theta_{i,f} = \measuredangle ({\bf k},{\bf p}_{i,f}) \sim 1 \: (\theta' \approx \theta_{i,f})
\end{equation}

\section{\label{sec:level1}The poles of the SB amplitude}

\begin{figure}[h!]
       \begin{center}
     \includegraphics[width=8cm]{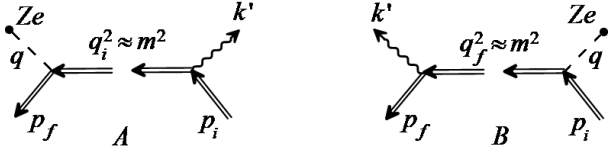}
     \end{center}
     \caption{Resonant spontaneous bremsstrahlung of an electron in the field of a nucleus and a plane electromagnetic wave.}
     \label{Fig:Figure 2}
  \end{figure}

The quasidiscrete structure of the system determines the resonant character of the amplitude \eqref{eq:13}-\eqref{eq:15}, \eqref{eq:31}, \eqref{eq:32}: electron + plane electromagnetic wave. As a result of the approximate fulfillment of the energy-momentum conservation laws, the 4-momentum of the intermediate electron in the scope of the developmental components of the effect emerges near mass surface.\\
The resonant conditions implement the following criterions \cite{20} for the channels A and B (see Fig. \ref{Fig:Figure 2}):
\begin{equation} \label{eq:36}
|\beta_j| \lesssim 1 \Rightarrow |q_j^2 - m^2| \lesssim {\frac {4 (k q_j)} {\omega \tau}},  \: j = i, f
\end{equation}

\begin{figure}[h!]
       \begin{center}
     \includegraphics[width=8cm]{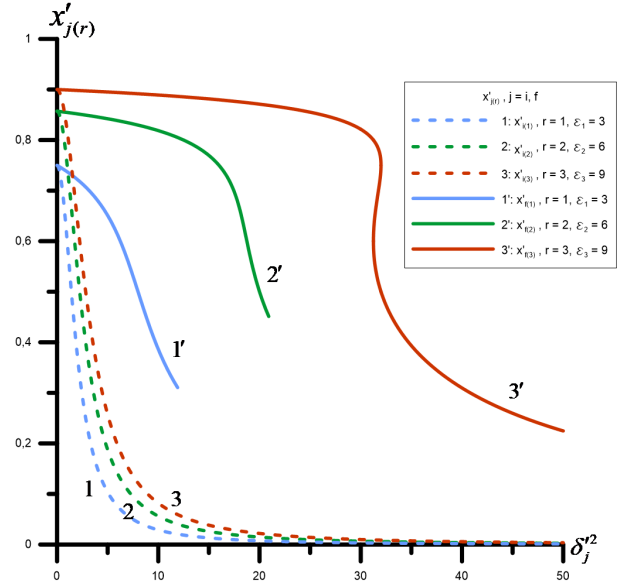}
     \end{center}
     \caption{The dependency of the resonant frequency of a spontaneous photon on the radiation angle. The dashed lines delineate the first three resonant frequencies of the channel A \eqref{eq:43} $(j = i)$, the solid lines illustrate the first three resonant frequencies of the channel B \eqref{eq:52}, \eqref{eq:48} $(j = f)$. The characteristic parameter $\varepsilon_1 = 3$ (the characteristic energy is $E_1 = 41.7 \: GeV$, the energy of the initial electron is $E_i = 125.1 \: GeV$).}
     \label{Fig:Figure 3}
  \end{figure}

The resonant parameters $\beta_{i,f}$ \eqref{eq:30} in the ambience of a weak field \eqref{eq:12} and kinematic conditions \eqref{eq:33}-\eqref{eq:35} represent a structure:
\begin{equation} \label{eq:37}
\beta_i = {\frac {r [\varepsilon_r - (1 + \varepsilon_r + \delta_i'^2) x']} {2 \varepsilon_r (1 - x')}} (\omega \tau)
\end{equation}
\begin{equation} \label{eq:38}
\beta_f = {\frac {r \{x' [1 + (1 - x')^2 \delta_f'^2] - \varepsilon_r (1 - x') \}} {2 \varepsilon_r (1 - x')}} (\omega \tau)
\end{equation}
Where
\begin{equation} \label{eq:39}
\delta_i'= {\frac {E_i \theta_i'} {m}}, \: \delta_f'= {\frac {E_i \theta_f'} {m}}, \: x' = {\frac {\omega'} {E_i}}
\end{equation}
\begin{equation} \label{eq:40}
\varepsilon_r = r \varepsilon_1, \: \varepsilon_1 = {\frac {E_i} {E_1}}, \: E_1 = {\frac {m^2} {4 \omega \sin^2(\theta_i/2)}}
\end{equation}
Here $r = 1, 2, 3,...$ - is the number of resonance (the amount of the photons of a wave that the electron absorbs in the laser-stimulated Compton-effect), $\varepsilon_r$ - the characteristic parameter of a process that is equal to the ratio of the initial electron energy to the characteristic energy $E_r = E_1/r$. The rest energy of an electron, the superposition of the energies of the absorbed photons of a wave and the angle between the momenta of the initial electron and propagating wave define the characteristic energy. Within the range of the optical frequencies the characteristic energy of the first resonance obtains the $E_1 \sim 10^2 GeV$ order of magnitude. With increase of the number of resonances the energy decreases \eqref{eq:40}. The works \cite{12,13,14,19,20} scrutinize the case of the moderate energies of the initial ultrarelativistic electrons $\varepsilon_1 \ll 1 \: (E_i \ll E_1)$ and the resonant frequency of the spontaneous photon was $\omega' \sim \varepsilon_1 E_i \ll E_i$. Additionally, previous investigations examined only the scattering of electrons at considerable angles. Moreover, the article \cite{47} evaluates the SB process of ultrarelativistic high-energy electrons specifically for the first resonance in the field of a plane monochromatic wave. The actual research analyzes the occurrence when $\varepsilon_r \gtrsim 1 \: (E_i \gtrsim E_r)$ for the several primary resonances within the field of a pulsed wave and when the initial electrons have considerable energies and scatter at small-scaled angles. With accordance to the resonant parameters \eqref{eq:37}, \eqref{eq:38} the resonant conditions \eqref{eq:36} for the channels A and B obtain a following form:
\begin{equation} \label{eq:41}
{\frac {r} {2 \varepsilon_r (1 - x')}} |\varepsilon_r - (1 + \varepsilon_r + \delta_i'^2) x'| \lesssim {\frac {1} {(\omega \tau)}} \ll 1,
\end{equation}
\begin{equation} \label{eq:42}
{\frac {r} {2 \varepsilon_r (1 - x')}} |x' [1 + \varepsilon_r + (1 - x')^2 \delta_f'^2] - \varepsilon_r| \lesssim {\frac {1} {(\omega \tau)}} \ll 1
\end{equation}
The equations \eqref{eq:41} and \eqref{eq:42} indicate that the expressions within the modulus brackets tend to zero. Therefore, the investigation derives the definitions of the resonant frequencies for the channels A and B. Thus, from the statement \eqref{eq:41} for the channel A:
\begin{equation} \label{eq:43}
x'_{i(r)}(\delta{'}^2_i) \approx {\frac {\varepsilon_i} {1 + \varepsilon_r + \delta_i'^2}}, \: x'_{i(r)}={\frac {\omega'_{i(r)}} {E_i}}
\end{equation}
This ratio designates that for the channel A the characteristic parameter $\varepsilon_r$ regulates the resonant frequency and that the resonant frequency formulates a single-value dependency on the angle of the spontaneous photon radiation in coordination to the initial electron momentum. In addition, when the spontaneous photon emits in the equal direction to the momentum of the initial electron $(\delta_i'^2 = 0)$ the resonant frequency reaches its maximum:
\begin{equation} \label{eq:44}
x'_{i(r)}(0) = x_{(r)}'^{max} = {\frac {\varepsilon_r}  {1+\varepsilon_r}}
\end{equation}
The resonant frequency decreases with the increase of the emission angle of the spontaneous photon and with $\delta_i'^2 \rightarrow \infty$ tends to zero (see dashed lines in Fig. \ref{Fig:Figure 3}). The frequency of a spontaneous photon increases with increase of the number of resonances.\\
From \eqref{eq:42} the research derives a cubic equation for the resonant frequency of the channel B:
\begin{equation} \label{eq:45}
\begin{aligned}
\delta_f'^2 x'^3_{f(r)} - 2\delta_f'^2 x'^2_{f(r)} + (1 + \delta_f'^2 + \varepsilon_r) x'_{f(r)} - \varepsilon_r \approx 0, \\  x'_{f(r)} = {\frac {\omega'_{f(r)}} {E_i}}
\end{aligned}
\end{equation}
The resonant frequency of the channel B depends on the characteristic parameter and on the angle of the spontaneous photon radiation in accordance to the momentum of a final electron. The study calculates the real roots of the equation \eqref{eq:45} within the interval $0 < x'_{f(r)} < 1$. The article \cite{47} scrutinized the expressions \eqref{eq:43} and \eqref{eq:45} for the first resonance appearance $(r = 1)$.\\
The statement \eqref{eq:45} has a single real root when the parameter value $\delta_f'^2 = 0$ and it corresponds to the scheme when photon radiates along the direction of the momentum of the final electron. Within the delineated conditions the resonant frequency of the spontaneous photon in channel B obtains maximal magnitude $x'_{f(r)} = x'^{max}_{(r)}$ that coincides with the coordinate value for the channel A (see \eqref{eq:44}). The consequential analysis represents the examination with limitation of $\delta_f'^2 \neq 0$. With reconsideration of \eqref{eq:45} (see \cite{47}) the investigation computes that within the range of the parameter $\delta_f'^2$:
\begin{equation} \label{eq:46}
\delta_{-(r)}'^2 < \delta_f'^2 < \delta_{+(r)}'^2
\end{equation}
Where
\begin{equation} \label{eq:47}
\begin{aligned}
\delta_{\pm (r)}'^2 = 3 ( 1 + \varepsilon_r ) + {\frac {1} {8}} (\varepsilon_r - 8) [(\varepsilon_i + 4) \pm \sqrt{\varepsilon_r (\varepsilon_r - 8)}], \\ \varepsilon_r > 8
\end{aligned}
\end{equation}
The resonant frequency of a spontaneous photon obtains three various possible magnitudes:
\begin{equation} \label{eq:48}
\begin{aligned}
x'_{f(r)1} = {\frac {2} {3}} + d'_{(r)} \cos({\frac {\varphi'_{(r)}} {3}}), \\ x'_{f(r)2} = {\frac {2} {3}} + d'_{(r)} \cos({\frac {\varphi'_{(r)}} {3}} +{\frac {2\pi} {3}}), \\ x'_{f(r)3} = {\frac {2} {3}} + d'_{(r)} \cos({\frac {\varphi'_{(r)}} {3}} + {\frac {4\pi} {3}})
\end{aligned}
\end{equation}
\begin{equation} \label{eq:49}
\begin{aligned}
d'_{(r)} = {\frac {2} {3 \delta'_f}} \sqrt{\delta_f'^2 - 3(1 + \varepsilon_r)}, \\ \cos \varphi'_{(r)} = {\frac {\delta'_f [9 \varepsilon_r - 2(9 + \delta_f'^2)]} {2[\delta_f'^2 - 3(1 + \varepsilon_r)]^{3/2}}}, \: 0 \leq \varphi'_{(r)} \leq \pi
\end{aligned}
\end{equation}
For the parameter $\delta_f'^2$ disposition within the area of:
\begin{equation} \label{eq:50}
0 < \delta_f'^2 \leq 3(1 + \varepsilon_r), \: if \: 0 < \varepsilon_r \leq 8
\end{equation}
\begin{equation} \label{eq:51}
0 < \delta_f'^2 \leq \delta_{-(r)}'^2, \: \delta_{+(r)}'^2 \leq \delta_f'^2 < \infty, \: if \: \varepsilon_r > 8
\end{equation}
The following equation distinctively defines the resonant frequency of a spontaneous photon
\begin{equation} \label{eq:52}
x'_{f(r)} =  {\frac {2} {3}} + (\alpha_{+(r)} + \alpha_{-(r)})
\end{equation}
Here
\begin{equation} \label{eq:53}
\alpha_{\pm (r)} =  [-{\frac {b_{(r)}} {2}} \pm \sqrt{Q_{(r)}}]^{1/3}, \: Q_{(r)} = ({\frac {a_{(r)}} {3}})^3 + ({\frac {b_{(r)}} {2}})^2
\end{equation}
\begin{equation} \label{eq:54}
\begin{aligned}
a_{(r)} = {\frac {1} {3 \delta_f'^2}} [3 (1 + \varepsilon_r) - \delta_f'^2], \: b_{(r)} = \\ = {\frac {1} {27 \delta_f'^2}} [2 (9 + \delta_f'^2) - 9 \varepsilon_r]
\end{aligned}
\end{equation}
It is important to underline that for the channel B there are two particular intervals of characteristic parameter values in which the distribution of the resonant frequencies and radiation angles of a spontaneous photon varies qualitatively. Therefore, for the range of $\varepsilon_r \leq 8$ \eqref{eq:50} the resonant frequency of a spontaneous photon constructs a single-value dependency on the angle of its emission with accordance to the momentum of a final electron. Additionally, the resonant frequency of a spontaneous photon fluctuates from maximum $x'^{max}_{(r)}$ \eqref{eq:44} when $\delta_f'^2 = 0$ to the minimum that is equal to:
\begin{equation} \label{eq:55}
x'^{min}_{f(r)} = {\frac {1} {3}} [2 - ({\frac {8 - \varepsilon_r} {1 + \varepsilon_r}})^{\frac {1} {3}}]
\end{equation}
when the parameter $\delta_{f(r)}'^2 = \delta_{f(r)max}'^2 = 3 (1 + \varepsilon_r)$ reaches maximum. For the radiation angles $\delta_f'^2 > \delta_{f(r)max}'^2$ the resonance of the channel B is absent (and an electron does not emit a spontaneous photon, see the solid lines on Fig. \ref{Fig:Figure 3} for $r = 1, 2$).\\
In the range of the characteristic parameter $\varepsilon_r > 8$ the dispersion of the resonant frequency of a spontaneous photon alterates substantially proportionally to the radiation angle coordinately to the momentum of a final electron (see the solid line on Fig. \ref{Fig:Figure 3} for $r = 3$). Thus, for the emission angles from the diapason:
\begin{equation} \label{eq:56}
0 \leq \delta_f'^2 < \delta_{-(r)}'^2
\end{equation}
the resonant frequency establishes a single-value dependency with a  radiation angle of a spontaneous photon and varies from the maximum $x'^{max}_{(r)}$ \eqref{eq:44} to the magnitude $x'_{f(r)-}$. For the emission angle $\delta_f'^2 = \delta_{-(r)}'^2$ the resonant frequency obtains two values. Within the range of the radiation angles:
\begin{equation} \label{eq:57}
\delta_{-(r)}'^2 < \delta_f'^2 < \delta_{+(r)}'^2
\end{equation}
the resonant frequency modifies from $x'_{f(r)-}$ to $x'_{f(r)+}$ and realizes three various magnitudes for each parameter $\delta_f'^2$ value. For the emission angle $\delta_f'^2 = \delta_{+(r)}'^2$ the resonant frequency attains two magnitudes. Finally, for the interval:
\begin{equation} \label{eq:58}
\delta_f'^2 > \delta_{+(r)}'^2
\end{equation}
Similar to the range \eqref{eq:56} the radiation angle of a spontaneous photon defines its resonant frequency with a single-value dependency and fluctuates from $x'_{f(r)+}$ to moderate magnitudes $x'_{f(r)} \ll 1$. research derives the resonant frequencies $x'_{f(r) \pm}$ with substitution $\delta_f'^2 = \delta_{\pm (r)}'^2$ in the solutions \eqref{eq:52}, \eqref{eq:53}.\\
In the non-resonant occurrence the frequency of a spontaneous photon and the energy of a final electron vary independently within the frameworks of the  approximate fulfillment of the energy conservation law. Moreover, the angles of emission of a spontaneous photon do not affect the energies of these particles. The resonant ambience indicates a fundamentally contrasting situation. Within the resonant conditions two statements determine the energies of a spontaneous photon and final electron: the energy conservation law and the resonant equation (see \eqref{eq:43} or \eqref{eq:45}). In addition, the angles of emission of a spontaneous photon in correlation to the momenta of the initial or final electrons define the possible energy spectrum of particles for channels A or B for various resonances. The number of resonance increases simultaneously with the frequency of the spontaneous photon.\\
It is important to underline that the chanels A and B do not interfiere as the angle of radiation of a sponteneous photon allocates resonant frequency in the reaction scheme A accordingly to the initial electron momentum and in the reaction scheme B appropriately to the final electron momentum. The analysis of the resonant frequencies of the channels A \eqref{eq:43} and B \eqref{eq:48}, \eqref{eq:52} (and Fig. \ref{Fig:Figure 3}) designates that within the structure of a single channel (A or B) the frequencies of a spontaneous photon are not equivalent for various resonances. Therefore, the interference of amplitudes for $r$ values is absent for each probable interaction channel (A or B).\\
Thus, the research proposes the exclusion from consideration of the interference of resonances within and between the channels A and B. Additionally, the article does not consider the scattering of an electron on a nucleus at zero angle of radiation $(\delta_i'^2 = \delta_f'^2 = 0)$.

\section{\label{sec:level1}The resonant cross-section of the SB process}

The consequential investigation simplifies the amplitude of the process \eqref{eq:13}-\eqref{eq:15}, \eqref{eq:31}-\eqref{eq:32} with accordance to the conditions of the resonant kinematics \eqref{eq:33}-\eqref{eq:34}, \eqref{eq:41}-\eqref{eq:42}. Therefore, the arguments of the Bessel functions \eqref{eq:24} in the amplitudes $M_{(l + r)}^0$ \eqref{eq:21} and $F_{(-r)}^{\mu}$ \eqref{eq:22} are small-scaled $(\gamma_{p'p} \lesssim \eta_0 \ll 1)$. Thus, the research establishes:
\begin{equation} \label{eq:59}
M_{(l + r)}^0 (p', p, \phi_j) \approx \tilde{\gamma}^0 L_0 (p', p, \phi_n) \approx \tilde{\gamma}^0 \: (l = -r)
\end{equation}
\begin{equation} \label{eq:60}
\begin{aligned}
F_{(-r)}^{\mu} (p', p, \phi_n) = \tilde{\gamma}^{\mu} L_{(-r)} (p', p, \phi_n) + \\ + L_{(-r + 1)} (p', p, \phi_n) \eta (\phi_1) [{\frac {m} {4 (k p')}} \hat{e}_{+} \hat{k} \tilde{\gamma}^{\mu} + {\frac {m} {4 (k p)}} \tilde{\gamma}^{\mu} \hat{k} \hat{e}_{+}]
\end{aligned}
\end{equation}
The following calculations are appropriate for the specifically selected type of the envelope function of the potential of the pulsed wave in the Gaussian form:
\begin{equation} \label{eq:61}
g (\phi_n) = e^{-(2 \phi_n)^2}, \: n = 1, 2
\end{equation}
With the implementation of the series representation of the Bessel functions in the amplitudes \eqref{eq:59}, \eqref{eq:60} on the small argument and after integration on the $d \phi_1$ for the channel A and on the $d \phi_2$ for the channel B the amplitude of the SB effect for the resonance number $r$ obtains the form:
\begin{equation} \label{eq:62}
S_{fi} = S_{(-r)} = {\frac {\pi^2 \tau^2 \omega Z e^3} {\sqrt{2 \omega' E_f E_i}}} {\frac {(\overline{u}_f B'_{(-r)} u_i)} {[{\bf q}^2 + q_0 (q_0 - 2 q_z)]}}
\end{equation}
Where
\begin{equation} \label{eq:63}
B'_{(-r)} = B'_{i(r)} + B'_{f(r)}
\end{equation}
Here $B'_{i(r)}$ and $B'_{f(r)}$ are the resonant amplitudes of the reaction schemes A and B.
\begin{equation} \label{eq:64}
B'_{i(r)} = U_{i(r)} \cdot [\tilde{\gamma}^0 (\hat{q}_i + m) (\varepsilon_{\mu}^{*} z_{i(r)}^{\mu})],
\end{equation}
\begin{equation} \label{eq:65}
B'_{f(r)} = U_{f(r)} \cdot [(\varepsilon_{\nu}^{*} z_{f(r)}^{\nu}) (\hat{q}_f + m) \tilde{\gamma}^0],
\end{equation}
\begin{equation} \label{eq:66}
U_{j(r)} = {\frac {\eta_0^r} {\sqrt{r} (k q_j)}} e^{-{\frac {\beta_j^2} {4 r}}} V_r (q_0, \beta_j),
\end{equation}
\begin{equation} \label{eq:67}
\begin{aligned}
V_r (q_0, \beta_j) = \int_{-\infty}^{\infty} d \phi \: e^{i (q_0 \tau + 2 \beta_j) \phi} {erf}(2 \sqrt{r} \phi + {\frac {i \beta_j} {2 \sqrt{r}}}), \\ j = i, f
\end{aligned}
\end{equation}
The sums on the $r$ are absent from the equations for the amplitudes \eqref{eq:64} and \eqref{eq:65} as a result of the absence of the interference of the resonances with various $r$. The expressions for $z_{i(r)}^{\mu}$, $z_{f(r)}^{\nu}$ organize in the following structure:
\begin{equation} \label{eq:68}
z_{i(r)}^{\mu} = \tilde{\gamma}^{\mu} c_i^r + c_i^{r - 1} [{\frac {m} {4 (k q_i)}} \hat{e}_{+} \hat{k} \tilde{\gamma}^{\mu} + {\frac {m} {4 (k p_i)}} \tilde{\gamma}^{\mu} \hat{k} \hat{e}_{+}],
\end{equation}
\begin{equation} \label{eq:69}
z_{f(r)}^{\nu} = \tilde{\gamma}^{\nu} c_f^r + c_f^{r - 1} [{\frac {m} {4 (k p_f)}} \hat{e}_{+} \hat{k} \tilde{\gamma}^{\nu} + {\frac {m} {4 (k q_f)}} \tilde{\gamma}^{\nu} \hat{k} \hat{e}_{+}],
\end{equation}
Where
\begin{equation} \label{eq:70}
c_i^r = {\frac {(-1)^r} {r!}} \gamma_i^r e^{i r \chi_{q_i p_i}}, \: c_f^r = {\frac {(-1)^r} {r!}} \gamma_f^r e^{i r \chi_{p_f q_f}}
\end{equation}
\begin{equation} \label{eq:71}
\gamma_j = r \sqrt{{\frac {u_j} {u_{jr}}} \cdot (1 - {\frac {u_j} {u_{jr}}})}, \: j = i, f
\end{equation}
In the equation \eqref{eq:71} $u_i, u_{ir}$ and $u_f, u_{fr}$ - are the relativistic-invariant parameters for the channels A and B.
\begin{equation} \label{eq:72}
u_i = {\frac {(k k')} {(k q_i)}}, \: u_{ir} = 2 r {\frac {(k p_i)} {m^2}}, \: u_f = {\frac {(k k')} {(k p_f)}}, \: u_{fr} = 2 r {\frac {(k q_f)} {m^2}}
\end{equation}
The fulfillment of the subsequential inequality for the number of the resonance and intensity of a wave regulates the applicability of the represented expansions of the Bessel functions:
\begin{equation} \label{eq:73}
{\frac {r} {\sqrt{r + 1}}} \ll \eta_0^{-1}
\end{equation}
The research derives the probability of the resonant process for the entire observation time in the absence of the interference of the channels A and B (the first and second addends in \eqref{eq:63}). Consequently, the investigation proposes:
\begin{equation} \label{eq:74}
\begin{aligned}
d w_{j(r)} = {\frac {\pi^4 \tau^4 \omega^2 (Z e^3)^2} {2 \omega' E_f E_i}} {\frac {|(\overline{u}_f B'_{j(r)} u_i)|^2} {[{\bf q}^2 + q_0 (q_0 - 2 q_z)]^2}} \cdot {\frac {d^3 p_f d^3 k'} {T (2 \pi)^6}}, \\ j = i, f
\end{aligned}
\end{equation}
Here $T$ - is a comparatively considerable $(T \gtrsim \tau)$ observation time interval. The study derives the differential cross-section of the SB process of an electron on a nucleus in the field of a pulsed light wave from the probability per unit of time by dividing it on the flux density of the scattering particles $\nu_i = |{\bf p}_i|/E_i$. The work scrutinizes the model for the unpolarized particles. Thus, the calculation defines the average value of the polarizations of the initial electrons and summation of the polarizations of the final electron and spontaneous photon with utilization of a standard procedure \cite{50}. The differential cross-section of the SB process for the channel A:
\begin{equation} \label{eq:75}
\begin{aligned}
d \sigma_{i(r)} = {\frac {\pi \tau (\omega \tau)^2} {16 (2 \pi)^3}} ({\frac {\tau} {T}}) {\frac {m^2 |{\bf q}_i|} {r (k q_i)^2}} e^{-{\frac {\beta_i^2} {2 r}}} d W'_{i(r)} (q_i, p_i) \cdot \\ \cdot {\frac {2 Z^2 r_e^2 m^2 [m^2 + E_f q_{i0} + {\bf p}_f {\bf q}_i]} {|{\bf q}_i| E_f [{\bf q}^2 + q_0 (q_0 - 2 q_z)]^2}} |V_r (q_0, \beta_i)|^2 d^3 p_f, \\ {\bf q} = {\bf p}_f - {\bf q}_i.
\end{aligned}
\end{equation}
Here $d W'_{i(r)} (q_i, p_i)$ - is the probability in the unit of time of radiation of a spontaneous photon with a 4-momentum $k'$ by an electron with a 4-momentum $p_i$ as a result of the absorption of $r$ - wave photons \cite{15} (the laser field-stimulated Compton-effect).
\begin{equation} \label{eq:76}
\begin{aligned}
d W'_{i(r)} (q_i, p_i) = {\frac {\alpha} {\omega' |{\bf p}_i|}} \eta_0^{2r} ({\frac {r^r} {r!}})^2 [{\frac {u_i} {u_{ir}}} (1 - {\frac {u_i} {u_{ir}}})]^{(r - 1)} \cdot \\ \cdot \{2 + {\frac {u_i^2} {(1 + u_i)}} - {\frac {4 u_i} {u_{ir}}} (1 - {\frac {u_i} {u_{ir}}}) \} d^3 k'
\end{aligned}
\end{equation}
The research arranges the integration in \eqref{eq:75} on the energy of the final electron with the substitution of the integration variable:
\begin{equation} \label{eq:77}
\zeta = {\frac {q_0} {\omega}} = {\frac {1} {\omega}} (E_f - E_i + \omega' - r \omega), \: d E_f = \omega d \zeta
\end{equation}
The magnitudes $\zeta$ from the following diapason affect fundamentally on the integral of \eqref{eq:75}:
\begin{equation} \label{eq:78}
|\zeta| \lesssim {\frac {1} {\omega \tau}} \ll 1
\end{equation}
Therefore, the investigation considers $\zeta = 0$ in the numerator of the equation \eqref{eq:75} with exception of the exponent addend $V_r (\omega \zeta, \beta_i)$ from \eqref{eq:67}. Additionally, the paper proposes the neglection of the corrections to the square of the transmitted momentum ${\bf q}^2$ in the denominator of \eqref{eq:75} as a result of the small-scaled magnitudes of the general approximations $|q_z| \omega \zeta \sim m^2 \omega \zeta / E_i \lesssim m^2 \omega / (E_i \omega \tau) \ll m^2 \omega / E_i$ to the value of the indicated momentum ${\bf q}^2 \sim m^4 / E_i^2 \lesssim m^2 \omega / E_i$ that produces the main impact on the resonant cross-section. Subsequently, the article accomplishes the integration in the \eqref{eq:75} on the $\zeta$. Finally, the resonant differential cross-sections of the SB process for the channels A and B obtain form:
\begin{equation} \label{eq:79}
d \sigma_{i(r)} = d \sigma_{(0)} (p_f, q_i) \cdot \Psi_{i(r)}^{res} \cdot d W'_{i(r)} (q_i, p_i)
\end{equation}
\begin{equation} \label{eq:80}
d \sigma_{f(r)} = d W'_{f(r)} (p_f, q_f) \cdot \Psi_{f(r)}^{res} \cdot d \sigma_{(0)} (q_f, p_i)
\end{equation}
Where the expression \eqref{eq:76} defines the $d W'_{i(r)} (q_i, p_i)$, the \eqref{eq:76}, \eqref{eq:72} construe the $d W'_{f(r)} (p_f, q_f)$ with replacement of the 4-momenta: $p_i \rightarrow q_f$, $q_i \rightarrow p_f$; $d \sigma_{(0)} (p_f, q_i)$ - is the differential cross-section of the scattering of an intermediate electron with 4-momentum $q_i$ on a nucleus without radiation (absorption) of wave photons \cite{11}:
\begin{equation} \label{eq:81}
\begin{aligned}
d \sigma_{(0)} (p_f, q_i) = 2 Z^2 r_e^2 {\frac {|{\bf p}_f| m^2} {|{\bf q}_i| {\bf q}^4}} [m^2 + E_f q_{i0} + {\bf p}_f {\bf q}_i] d \Omega_f, \\ {\bf q} = {\bf p}_f - {\bf q}_i
\end{aligned}
\end{equation}
The $d \sigma_{(0)} (q_f, p_i)$ derives from \eqref{eq:81} with substitution of the 4-momenta: $q_i \rightarrow p_i$, $p_f \rightarrow q_f$. The resonant functions $\Psi_{i(r)}^{res}$ and $\Psi_{f(r)}^{res}$ have structure:
\begin{equation} \label{eq:82}
\begin{aligned}
\Psi_{i(r)}^{res} = {\frac {m^2 |{\bf q}_i|} {64 \pi^2}} {\frac {(\omega \tau)^2 P_{(r)}^{res} (\beta_i)} {r (k q_i)^2}}, \\ \Psi_{f(r)}^{res} = {\frac {m^2 |{\bf p}_f|} {64 \pi^2}} {\frac {(\omega \tau)^2 P_{(r)}^{res} (\beta_f)} {r (k q_f)^2}}
\end{aligned}
\end{equation}
where the resonant parameter $\beta_j$ defines the function of the resonance profile $P_{(r)}^{res} (\beta_j)$:
\begin{equation} \label{eq:83}
\begin{aligned}
P_{(r)}^{res} (\beta_j) = e^{(-{\frac {\beta_j^2} {2 r}})} {\frac {1} {2 \rho}} \int_{-\rho}^{\rho} |{erf} (2 \sqrt{r} \phi + {\frac {i \beta_j} {2 \sqrt{r}}})|^2 d \phi, \\ j = i, f; \: \rho = T / \tau
\end{aligned}
\end{equation}

\begin{figure}[h!]
       \begin{center}
     \includegraphics[width=8cm]{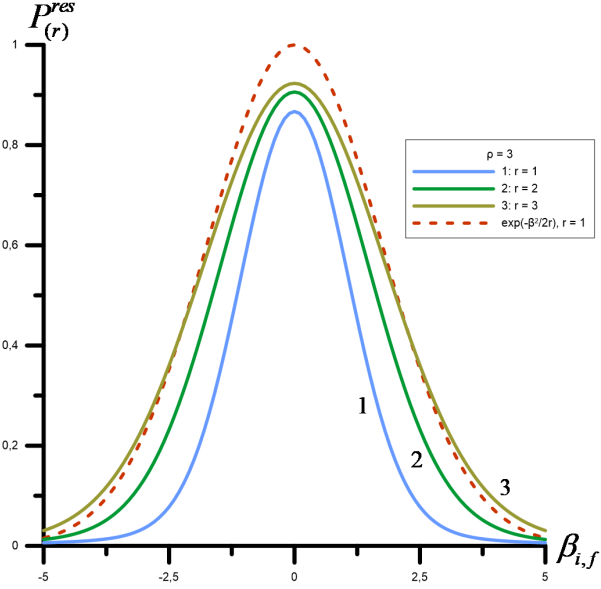}
     \end{center}
     \caption{The dependency of the function of the resonance profile $P_{(r)}^{res}$ \eqref{eq:83} on the parameter $\beta_{i, f}$ \eqref{eq:30}, \eqref{eq:12} when $\rho = 3$. The lines $1$, $2$, $3$ designate the first, second and third resonances. The dashed line illustrates the Gauss function.}
     \label{Fig:Figure 4}
  \end{figure}

The study analyzes the represented function with the condition of the resonant parameter value of $|\beta_j| \ll 1$. In order to simplify the calculations in the delineated approximation the function $P_{(r)}^{res} (\beta_j)$ of the resonance profile transforms into the Breit-Wigner equation \cite{20}:
\begin{equation} \label{eq:84}
P_{(r)}^{res} (|\beta_j| \ll 1) \approx {\frac {a_{(r)} \Lambda_{j(r)}^2} {[(q_j^2 - m^2)^2 + \Lambda_{j(r)}^2]}} \approx P_{max(r)}^{res} = a_{(r)}
\end{equation}
Here $\Lambda_{j(r)}$ - is the width of the according resonance:
\begin{equation} \label{eq:85}
\Lambda_{j(r)} = {\frac {4 r c_{(r)} (k q_j)} {(\omega \tau)}}, \: j = i, f; \: c_{(r)} = \sqrt{{\frac {2 a_{(r)}} {r (a_{(r)} - 2 r b_{(r)})}}}
\end{equation}
\begin{equation} \label{eq:86}
a_{(r)} = {\frac {1} {\rho}} \int_{0}^{\rho} {erf}^2 (2 \sqrt{r} \phi) d \phi
\end{equation}
\begin{equation} \label{eq:87}
\begin{aligned}
b_{(r)} = {\frac {1} {2 \rho \sqrt{\pi r}}} [{\frac {1} {\sqrt{2 \pi} r}} {erf} (2 \sqrt{2 r} \rho) + \\ + 2 \int_{0}^{\rho} \phi \: {erf} (2 \sqrt{r} \phi) \: e^{-4 r \phi^2} d \phi]
\end{aligned}
\end{equation}
The parameters for the first three resonances are $a_{(r)} \sim 1$, $c_{(r)} \sim 1$, $b_{(r)} \sim 10^{-2}$. The expression \eqref{eq:85} indicates that the parameter $a_{(r)}$ \eqref{eq:86} that is dependant on the number of resonance and on the ratio of the observation time to the momentum (criterion $\rho$ \eqref{eq:83}) establishes the maximum of the function of the resonance profile. The Fig. \ref{Fig:Figure 4} outlines the function of the resonance profile \eqref{eq:83} for the first, second, and third resonances. The demonstrated curves specify that the resonances occur when the resonant parameters are $|\beta_{i, f}| \lesssim 1$. With $|\beta_{i, f}| \gg 1$ the resonance profile function exponentially decreases.

\section{\label{sec:level1}The analysis of the resonant differential cross-section of the SB process of ultrarelativistic electrons}

\begin{figure}[h!]
   \begin{minipage}{.5\textwidth}
       \includegraphics[width=7cm]{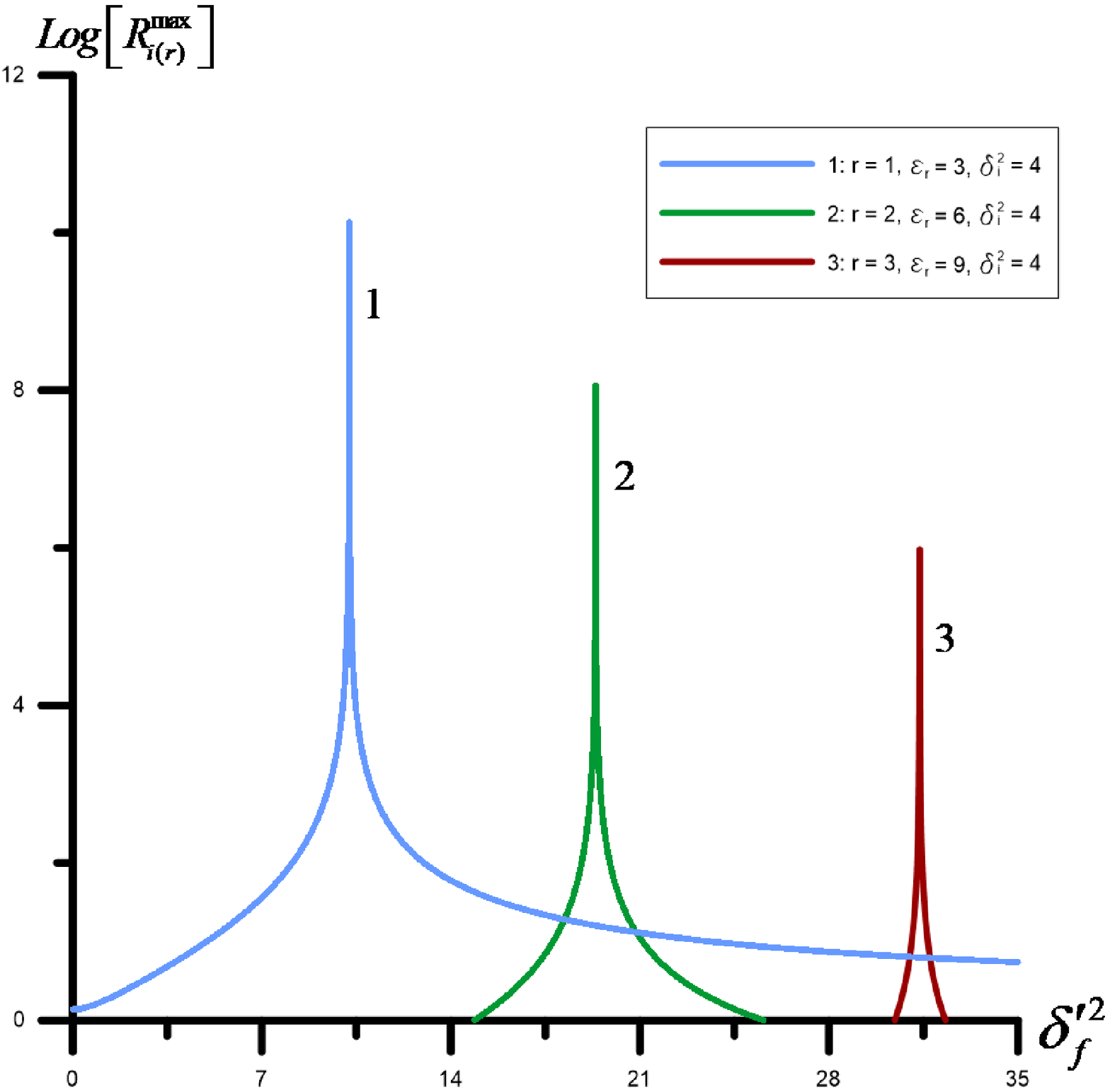}
  \end{minipage}
  \begin{minipage}{.5\textwidth}
      \includegraphics[width=7cm]{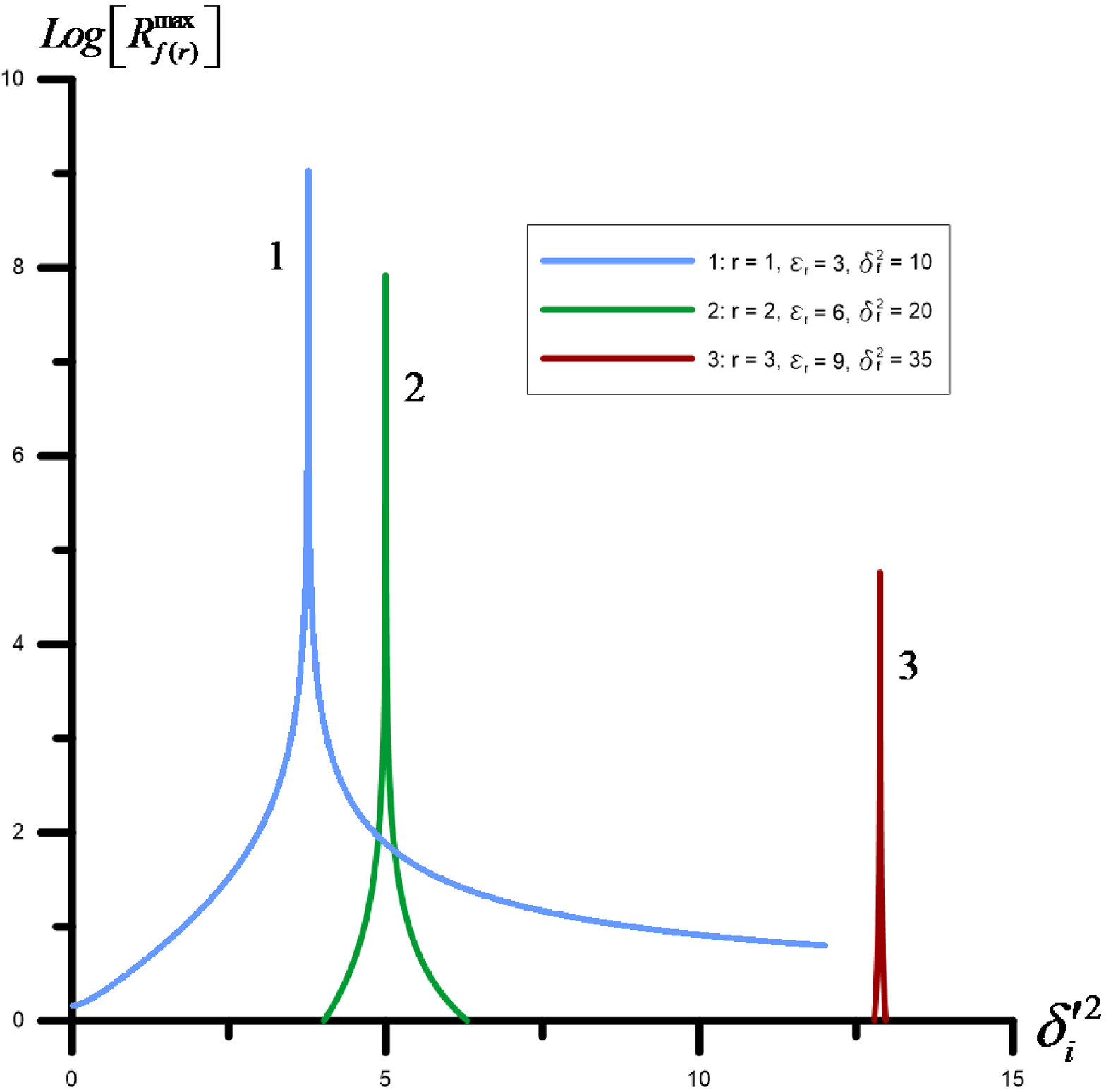}
  \end{minipage}
  \caption{The dependency of the functions $R^{max}_{i(r)}$ and $R^{max}_{f(r)}$ on the parameters $\delta'^2_f$ (for the channel A) and $\delta'^2_i$ (for the channel B) for the fixed magnitude of the spontaneous photon frequency. The plot illustrates the first, second and third resonances of the \eqref{eq:113} equation. The characteristic parameter is $\varepsilon_1 = 3$ (the characteristic energy is $E_1 = 41.7 \: GeV$, the energy of the initial electron is $E_i = 125.1 \: GeV$).}
  \label{Fig:Figure 5}
\end{figure}

The research appoints the transformation of the resonant differential cross-section of the SB effect \eqref{eq:79}-\eqref{eq:83} under the conditions of the resonant kinematics \eqref{eq:33}-\eqref{eq:35}. The calculations derive:
\begin{equation} \label{eq:88}
\begin{aligned}
d\sigma_{j(r)} = \alpha r_e^2 Z^2 \eta_0^{2 r} {\frac {r (\omega \tau)^2 (1 - x'_{j(r)})} {8 \pi d_{j(r)}^2}} \cdot {\frac {K_{j(r)}} {\varepsilon_r^2}} P_{(r)}^{res} (\beta_j) \cdot \\ \cdot x'_{j(r)} d x'_{j(r)} d \delta_i'^2 d \delta_f'^2 d\varphi'_{-}, \: j = i, f
\end{aligned}
\end{equation}
Where $\varphi'_{-}$ - is the angle between the planes $({\bf k}',{\bf p}_i)$ and $({\bf k}',{\bf p}_f)$. The equation \eqref{eq:83} and the resonant parameters from the statements \eqref{eq:37}, \eqref{eq:38} define the function of the resonant profile $P_{(r)}^{res} (\beta_j)$. The probability of the laser-stimulated Compton-effect characterizes the functions $K_{j(r)}$ that are equal to:
\begin{equation} \label{eq:89}
\begin{aligned}
K_{j(r)} = ({\frac {r^r} {r!}})^2 [{\frac {x'_{j(r)}} {\varepsilon_r (1 - x'_{j(r)})}} \cdot (1 - {\frac {x'_{j(r)}} {\varepsilon_r (1 - x'_{j(r)})}})]^{(r - 1)} \times \\ \times \{2 + {\frac {x_{j(r)}'^2} {(1 - x'_{j(r)})}} - {\frac {4 x'_{j(r)}} {\varepsilon_r (1 - x'_{j(r)})}} (1 - {\frac {x'_{j(r)}} {\varepsilon_r (1 - x'_{j(r)})}})\}
\end{aligned}
\end{equation}
The square of the transmitted momentum regulates the $d_{j(r)}$ magnitudes.
\begin{equation} \label{eq:90}
\begin{aligned}
d_{j(r)} = d_0 (x'_{j(r)}) + ({\frac {m} {2 E_i}})^2 [g_0^2 (x'_{j(r)}) + \\ + {\frac {\varepsilon_r} {\sin^2 (\theta_i / 2)}} (\varepsilon_r + g_0 (x'_{j(r)}))]
\end{aligned}
\end{equation}
Where
\begin{equation} \label{eq:91}
\begin{aligned}
d_0 (x'_{j(r)}) = \tilde{\delta}'^2_f + \delta_i^2 - 2 \delta_i \tilde{\delta}'_f \cos (\varphi_{-}), \: g_0 (x'_{j(r)}) = \\ = (1 + \delta_i^2) - {\frac {(1 + \tilde{\delta}'^2_f)} {(1 - x'_{j(r)})}}
\end{aligned}
\end{equation}
\begin{equation} \label{eq:92}
\tilde{\delta}'_f = (1 - x'_{j(r)}) \delta'_f
\end{equation}
The differential cross-section of the SB process in the absence of the external laser field $d \sigma_{*}$ has the following form within the equivalent frequency range \cite{47,50}:
\begin{equation} \label{eq:93}
\begin{aligned}
d\sigma_{*} = {\frac {2} {\pi}} Z^2 \alpha r_e^2 (1 - x')^3 {\frac {[D_0 (x') + (m / E_i)^2 D_1 (x')]} {[d_0 (x') + (m / 2 E_i)^2 g_0^2 (x')]^2}} \cdot \\ \cdot {\frac {d x'} {x'}} d \delta_i'^2 d \delta_f'^2 d\varphi'_{-}
\end{aligned}
\end{equation}
Here:
\begin{equation} \label{eq:94}
\begin{aligned}
D_0(x') = {\frac {\delta'^2_i} {(1 + \delta'^2_i)^2}} + {\frac {\tilde{\delta}'^2_f} {(1 + \tilde{\delta}'^2_f)^2}} + {\frac {x{'}^2} {2 (1 - x')}} \cdot \\ \cdot {\frac {(\delta'^2_i + \tilde{\delta}'^2_f)} {(1 + \delta'^2_i)(1 + \tilde{\delta}'^2_f)}} - [(1 - x') + {\frac {1} {(1 - x')}}] \cdot \\ \cdot {\frac {\delta'_i \tilde{\delta}'_f} {(1 + \delta'^2_i)(1 + \tilde{\delta}'^2_f)}} \cos \varphi_{-}
\end{aligned}
\end{equation}
\begin{equation} \label{eq:95}
D_1(x') = b_i(x') + {\frac {b_f(x')} {(1 - x')^2}}
\end{equation}
\begin{equation} \label{eq:96}
b_i(x') =  {\frac {\delta'^2_i} {12 (1 + \delta'^2_i)^3}} \xi_i
\end{equation}
\begin{equation} \label{eq:97}
\begin{aligned}
\xi_i = [(1 - x') + {\frac {1} {(1 - x')}}] (9 + 4\delta'^2_i +3\delta'^4_i) - \\ - 2(1 - \delta'^2_i)(3 - \delta'^2_i) - {\frac {x'^2} {(1 - x')}} (9 + 2 \delta'^2_i + \delta'^4_i)
\end{aligned}
\end{equation}
With substitution $\delta'^2_i \rightarrow \tilde{\delta}'^2_f$ in the expressions \eqref{eq:96}, \eqref{eq:97} the study acquires $b_f(x')$. It is important to underline that the work implements small-scaled corrections proportional to $\sim (m / E_i)^2 \ll 1$   in the differential cross-sections \eqref{eq:88}, \eqref{eq:90} and \eqref{eq:93}. Moreover, the indicated corrections apportion a dominant impact in the coordinate differential cross-section when $\delta_i' \sim \tilde{\delta}'_f \gtrsim 1$ and
\begin{equation} \label{eq:98}
\varphi_{-} \lesssim {\frac {m} {E_i}}, \: |\delta_i' - \tilde{\delta}'_f| \lesssim {\frac {m} {E_i}}
\end{equation}
Within this limitations the values $D_0 (x') \rightarrow 0$, $d_0 (x') \rightarrow 0$ and the according differential cross-sections realize a sharp maximum (see Fig. \ref{Fig:Figure 5}). Thus, the differential cross-sections in the absence of the field \eqref{eq:93} and in the field of a wave \eqref{eq:88} in the kinematical diapason \eqref{eq:98} obtain the following order of magnitude:
\begin{equation} \label{eq:99}
d\sigma_{*} \sim Z^2 \alpha r_e^2 ({\frac {E_i} {m}})^2, \: d\sigma_{j(r)} \lesssim Z^2 \alpha r_e^2 ({\frac {E_i} {m}})^4 (\eta_0^r \cdot \omega \tau)^2
\end{equation}
After integration of the resonant differential cross-sections \eqref{eq:88} and cross-section without external field \eqref{eq:93} on the azimuthal angle $\varphi_{-}$ the study derives:
\begin{equation} \label{eq:100}
\begin{aligned}
d\sigma_{j(r)} = \alpha r_e^2 Z^2 \eta_0^{2 r} {\frac {r (\omega \tau)^2 (\tilde{\delta}'^2_f + \delta'^2_i) (1 - x'_{j(r)})} {4 G_{j(r)}^{3/2} \cdot \varepsilon_r^2}} \cdot \\ \cdot K_{j(r)} P_{(r)}^{res} (\beta_j) \cdot d x'_{j(r)} d \delta_i'^2 d \delta_f'^2
\end{aligned}
\end{equation}
where
\begin{equation} \label{eq:101}
\begin{aligned}
G_{j(r)} = (\tilde{\delta}'^2_f - \delta'^2_i)^2 + {\frac {1} {2}} ({\frac {m} {E_i}})^2 (\tilde{\delta}'^2_f + \delta'^2_i) [g_0^2 (x'_{j(r)}) + \\ + {\frac {\varepsilon_r} {\sin^2 (\theta_i / 2)}} (\varepsilon_r + g_0(x'_{j(r)}))]
\end{aligned}
\end{equation}
In the represented conditions the differential cross-section of the SB process without the laser field:
\begin{equation} \label{eq:102}
d\sigma_{*} = 4 Z^2 \alpha r_e^2 {\frac {(1 - x')^3} {G_{0}^{3/2}}} (\delta'^2_i + \tilde{\delta}'^2_f) \cdot D_2 (x') \cdot {\frac {d x'} {x'}} d \delta_i'^2 d \delta_f'^2
\end{equation}
The article acquires $G_{(0)}$ from the equation for $G_{j(r)}$ \eqref{eq:101} with prerequisite $\varepsilon_r = 0$. The supplementary addends:
\begin{equation} \label{eq:103}
D_2(x') = D'_0(x') + (m / E_i)^2 D'_1(x')
\end{equation}
\begin{equation} \label{eq:104}
\begin{aligned}
D'_0(x') = {\frac {\delta'^2_i} {(1 + \delta'^2_i)^2}} + {\frac {\tilde{\delta}'^2_f} {(1 + \tilde{\delta}'^2_f)^2}} + {\frac {x{'}^2} {2 (1 - x')}} \cdot \\ \cdot {\frac {(\delta'^2_i + \tilde{\delta}'^2_f)} {(1 + \delta'^2_i)(1 + \tilde{\delta}'^2_f)}} - [(1 - x') + {\frac {1} {(1 - x')}}] \cdot \\ \cdot {\frac {2 \delta'^2_i \tilde{\delta}'^2_f} {(1 + \delta'^2_i)(1 + \tilde{\delta}'^2_f)(\delta'^2_i + \tilde{\delta}'^2_f)}}
\end{aligned}
\end{equation}
\begin{equation} \label{eq:105}
\begin{aligned}
D_1' (x') = D_1 (x')+ [(1 - x') + {\frac {1} {(1 - x')}}] \cdot \\ \cdot {\frac {g_0^2 (x') \delta'^2_i \tilde{\delta}'^2_f} {2 (1 + \delta'^2_i) (1 + \tilde{\delta}'^2_f) (\delta'^2_i + \tilde{\delta}'^2_f)^2}}
\end{aligned}
\end{equation}
The investigation reexamines the expression for the function of the resonance profile $P_{(r)}^{res} (\beta_j)$ in \eqref{eq:88} when $\beta_j \ll 1$:
\begin{equation} \label{eq:106}
\begin{aligned}
P_{(r)}^{res} (\beta_j \ll 1) \approx {\frac {a_{(r)} \Gamma_{j(r)}^2} {[(\delta'^2_j - \delta'^2_{j(r)})^2 + \Gamma_{j(r)}^2]}} \approx \\ \approx P_{max(r)}^{res} = a_{(r)}, \: j = i, f
\end{aligned}
\end{equation}
Here the parameters $\delta'^2_{j(r)}$ fulfill the corresponding equations for the resonant frequencies (for the channel A - \eqref{eq:43}, for the channel B - \eqref{eq:45}). In addition, the transit width for the channels A and B:
\begin{equation} \label{eq:107}
\begin{aligned}
\Gamma_{i(r)} = {\frac {2 \varepsilon_r c_{(r)} (1 - x'_{i(r)})} {(\omega \tau) x'_{f(r)}}}, \:  \Gamma_{f(r)} = {\frac {2 \varepsilon_r c_{(r)}} {(\omega \tau) x'_{f(r)} (1 - x'_{f(r)})}}
\end{aligned}
\end{equation}
It is important to underline that the calculations for the function of the resonance profile \eqref{eq:83}, \eqref{eq:106} are correct with the restriction that the transit width significantly exceeds the radiation width
\begin{equation} \label{eq:108}
\Gamma_{j(r)} \gg \Upsilon_j
\end{equation}
where $\Upsilon_j$ - is the radiation width of the resonance \cite{47}:
\begin{equation} \label{eq:109}
\Upsilon_i = {\frac {1} {4}} \alpha \eta^2 K_1 ({\frac {1 - x'_i} {x'_i}}), \: \Upsilon_f = {\frac {1} {4}} \alpha \eta^2 {\frac {K_1} {x'_f (1 - x'_f)}}
\end{equation}
\begin{equation} \label{eq:110}
K_1 = (1 - {\frac {4} {\varepsilon_1}} - {\frac {8} {\varepsilon_1^2}}) \ln (1 + \varepsilon_1) + {\frac {1} {2}} + {\frac {8} {\varepsilon_1}} - {\frac {1} {2 (1 + \varepsilon_1)^2}}
\end{equation}
Accounting the \eqref{eq:107}, \eqref{eq:109} the condition \eqref{eq:108} generates an upper limit of the duration time of the laser pulse:
\begin{equation} \label{eq:111}
\omega \tau \ll 10^3 \eta_0^{-2}
\end{equation}
For example, $\eta_0 = 10^{-1}$ from \eqref{eq:111} the order is $\omega \tau \ll 10^5$. The regulations \eqref{eq:3} and \eqref{eq:111} define the intervals of the possible laser pulse duration that correlate to the derived resonant cross-sections.\\
The research scrutinizes the ratio of the maximal resonant differential cross-section $d \sigma_{j(r)}^{max} \approx \sigma_{j(r)} (|\beta_j| \ll 1)$ \eqref{eq:100}, \eqref{eq:106} to the according differential cross-section in the absence of the light field \eqref{eq:102}:
\begin{equation} \label{eq:112}
\begin{aligned}
R^{max}_{j(r)} = {\frac {d\sigma^{max}_{j(r)}} {d\sigma_{*}}} = r a_{(r)} ({\frac {\eta_0^r \omega \tau} {4 \varepsilon_r}})^2 ({\frac {x'_{j(r)}} {1 - x'_{j(r)}}})^2 \cdot \\ \cdot {\frac {K_{j(r)}} {D_2 (x'_{j(r)})}} [{\frac {G_{0}} {G_{j(r)}}}]^{3/2}, \: j = i, f
\end{aligned}
\end{equation}
The expression \eqref{eq:112} designates the magnitudes of the resonant differential cross-section of the SB effect (in the units of the coordinate differential cross-section of the SB without the laser field) for the channels A and B with  simultaneous registration of the radiation angles of the final electron and spontaneous photon (parameters $\delta'^2_i$ and $\delta'^2_f$) and the frequency of the spontaneous photon in the range from $\omega'_{i(r)}$ to $[\omega'_{i(r)} + d \omega'_{i(r)}]$ (for channel A) and from $\omega'_{f(r)}$ to $[\omega'_{f(r)} + d \omega'_{f(r)}]$ (for channel B). The emission angle of a spontaneous photon in correlation to the momentum of the initial electron (parameter $\delta'^2_i$) delineates the resonant frequency $\omega'_{i(r)}$ and the energy of a final electron $E_f \approx E_i - \omega'_{i(r)}$ for the channel A. However, the specified values do not depend on the angle of radiation of the final electron (parameter $\delta'^2_f$). In contrast, for the channel B the emission angle of a spontaneous photon in accordance to the momentum of the final electron (parameter $\delta'^2_f$) allocates the resonant frequency $\omega'_{f(r)}$ and the energy of a final electron $E_f \approx E_i - \omega'_{f(r)}$. The indicated magnitudes do not depend on the parameter $\delta'^2_i$.

\begin{figure}[h!]
   \begin{minipage}{.5\textwidth}
       \includegraphics[width=7cm]{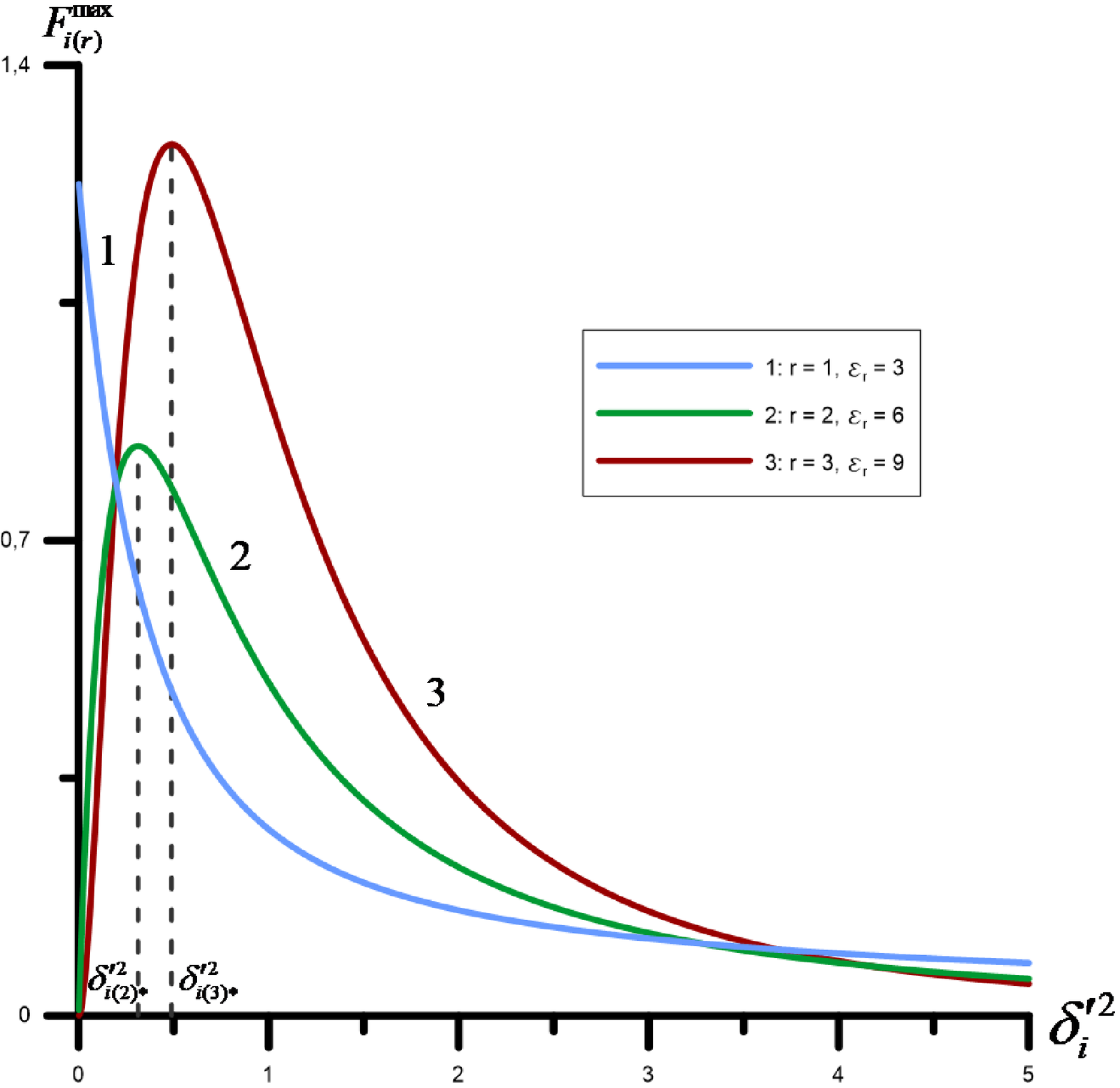}
  \end{minipage}
  \begin{minipage}{.5\textwidth}
      \includegraphics[width=7cm]{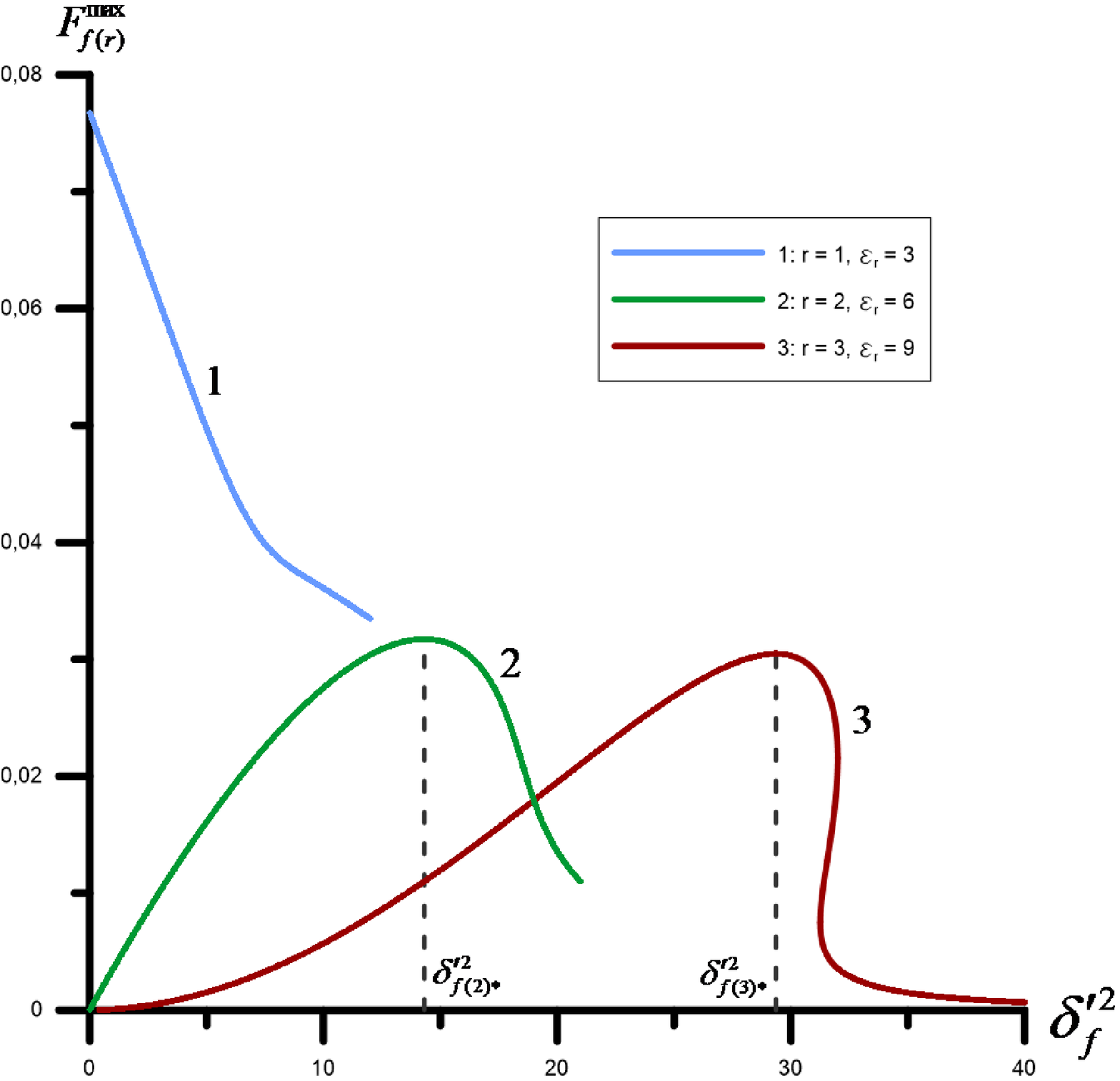}
  \end{minipage}
  \caption{The dependencies of the functions $F_{i(r)}^{max}$ (for the channel A) and $F_{f(r)}^{max}$ (for the channel B) \eqref{eq:120} on the corresponding parameters $\delta_i'^2$ and $\delta_f'^2$ for the first, second and third resonances. The characteristic parameter is $\varepsilon_1 = 3$ (the characteristic energy is  $E_1 = 41.7 \: GeV$, the energy of the initial electron is $E_i = 125.1 \: GeV$).}
  \label{Fig:Figure 6}
\end{figure}

The Fig. \ref{Fig:Figure 5} outlines the graphs of the functions $R^{max}_{j(r)}$ on the coordinate parameters $\delta'^2_j$ for the channels A and B for the fixed value of the frequency of a spontaneous photon. The Fig. \ref{Fig:Figure 5} indicates that for the occurrence when the parameters $\delta'^2_f = \delta'^2_i / (1 - x'_{i(r)})^2$ (for the channel A) and $\delta'^2_i = (1 - x'_{f(r)})^2 \delta'^2_f$ (for the channel B) coincide the dependencies realize sharp maximums that associate with small-scaled transmitted momenta (the corrections $\sim (m / E_i)^2 \ll 1$ that produce the general impact on the cross-section \eqref{eq:100}, \eqref{eq:102}, \eqref{eq:112}). With increase of the number of resonances (amount of photons that the electron absorbs) the level of the maximum peaks decreases essentially.\\
Subsequently, the research arranges integration of the resonant differential cross-section on the parameter $\delta'^2_f$ (for the channel A) and on the parameter $\delta'^2_i$ (for the channel B) \eqref{eq:100} for the fixed values of the frequency of a spontaneous photon (see Fig. \ref{Fig:Figure 5}):
\begin{equation} \label{eq:113}
d \sigma'_{i(r)} = (\alpha r_e^2 Z^2) g_r F_{i(r)} d x'_{i(r)} d \delta_i'^2
\end{equation}
\begin{equation} \label{eq:114}
d \sigma'_{f(r)} = (\alpha r_e^2 Z^2) g_r F_{f(r)} d x'_{f(r)} d \delta_f'^2
\end{equation}
Where
\begin{equation} \label{eq:115}
g_r = ({\frac {E_1} {r m}})^2 (\eta_0^r \omega \tau)^2
\end{equation}
\begin{equation} \label{eq:116}
F_{i(r)} = {\frac {r} {\varepsilon_r^2}} {\frac {x'_{i(r)}} {(1 - x'_{i(r)})}} K_{i(r)} P_{(r)}^{res} (\beta_i)
\end{equation}
\begin{equation} \label{eq:117}
F_{f(r)} = {\frac {r} {\varepsilon_r^2}} x'_{f(r)} (1 - x'_{f(r)}) K_{f(r)} P_{(r)}^{res} (\beta_f)
\end{equation}
The expression \eqref{eq:113} defines the resonant differential cross-section of the SB process for the channel A with simultaneous registration of the frequency and radiation angle of a spontaneous photon in accordance to the initial electron (without dependency on the angle of emission of the final electron). The equation \eqref{eq:114} characterizes the resonant resonant differential cross-section of the SB effect for the channel B with concurrent evaluation of the frequency and radiation angle of a spontaneous photon in correlation to the final electron. Additionally, the function $g_r$ \eqref{eq:115} is equivalent for the both reaction channels. The couplet of factors (the first is considerable and coordinates with the small-scaled transmitted momenta and the characteristic energy of the process \eqref{eq:40}, the transit width of the resonances establishes the second factor) specifies the represented value. Therefore, for $E_1 = 41.7 GeV$, $\eta_0 = 10^{-1} (= 10^{-2})$, $(\omega \tau) = 10^2$ from \eqref{eq:115} the investigation obtains: for the first resonance $g_1 \approx 0.70 \cdot 10^{12} (\approx 0.70 \cdot 10^{10})$, for the second resonance $g_2 \approx 1.75 \cdot 10^9 (\approx 1.75 \cdot 10^5)$, and for the third resonance $g_3 \approx 0.78 \cdot 10^7 (\approx 7.8)$. The function $g_r$ substantially depends on the intensity of a wave and on the resonance number $r$. With increase of the amount of the absorbed photons $r$ and decrease of the intensity of a wave the laser field-stimulated Compton-effect becomes suppressed. The functions $F_{i(r)}$ \eqref{eq:116} and $F_{f(r)}$ \eqref{eq:117} determine the quantitative level of the resonant differential cross-sections for the channel A \eqref{eq:113} and B \eqref{eq:114} for each resonance (classified with value $r$). Consequently, with the conditions $|\beta_i| \ll 1$ and $|\beta_f| \ll 1$ the indicated functions and the coordinate resonant cross-sections attain their maximal magnitudes:
\begin{equation} \label{eq:118}
\begin{aligned}
d \sigma'^{max}_{i(r)} = \alpha r_e^2 Z^2 \Phi_{i(r)}^{max} d x'_{i(r)} d \delta_i'^2, \\ d \sigma'^{max}_{f(r)} = \alpha r_e^2 Z^2 \Phi_{f(r)}^{max} d x'_{f(r)} d \delta_f'^2
\end{aligned}
\end{equation}
\begin{equation} \label{eq:119}
\Phi_{i(r)}^{max} = g_r F_{i(r)}^{max}, \: \Phi_{f(r)}^{max} = g_r F_{f(r)}^{max}
\end{equation}
\begin{equation} \label{eq:120}
\begin{aligned}
F_{i(r)}^{max} = {\frac {r a_{(r)}} {\varepsilon_r^2}} {\frac {x'_{i(r)}} {(1 - x'_{i(r)})}} K_{i(r)}, \\ F_{f(r)}^{max} = {\frac {r a_{(r)}} {\varepsilon_r^2}} x'_{f(r)} (1 - x'_{f(r)}) K_{f(r)},
\end{aligned}
\end{equation}
The Fig. \ref{Fig:Figure 6} illustrates the functions \eqref{eq:120}. The Table 1 delineates the values of the resonant frequencies and the according differential cross-sections (functions $\Phi_{j(r)}^{max}$) for the channels A and B for the prerequisites of a couple of possible intensities of a wave for the first, second and third resonances when $E_1 = 41.7 \: GeV$, $E_i = 125.1 \: GeV$, $(\omega \tau) = 10^2$. The plots \ref{Fig:Figure 6}a and \ref{Fig:Figure 6}b designate that within the whole range of the parameters $\delta_i'^2$ and $\delta_f'^2$ variation the function $F_{i(r)}^{max}$ exceeds the $F_{f(r)}^{max}$ for approximately two orders of the magnitude. Therefore, the radiation of a spontaneous photon within the channel B is diminutive in comparison to the channel A. Under the resonant conditions spontaneous photons radiate through the channel A and the resonant cross-section for the first resonance can reach the order of $\Phi_{i(1)}^{max} \sim 10^{12}$ when $\eta_0 = 0.1$. With decrease of the wave intensity the represented value decreases as the power factor $\sim \eta_0^{2 r}$.

\begin{table*}[]
\centering
\begin{tabular}{|c|c|c|c|c|c|}
\hline
\multicolumn{6}{|c|}{Table 1}           \\ \hline
                  r & Channel & $\delta_j'^2$ & $\omega'_{j(r)}, GeV$ & $\Phi_{j(r)}^{max} (\eta_0 = 0.1)$ & $\Phi_{j(r)}^{max} (\eta_0 = 0.01)$ \\ \hline
\multirow{2}{*}{} 1 & A & $\delta_i'^2 \geq 0$ & $\omega'_{i(1)} \leq 93.823$ & $\Phi_{i(1)}^{max} \lesssim 10^{12}$ & $\Phi_{i(1)}^{max} \lesssim 10^{10}$ \\ \cline{2-6}
                  & B & $0 \leq \delta_f'^2 \leq 12$ & $38.480 \leq \omega'_{f(1)} \leq 93.823$ & $\Phi_{f(1)}^{max} \sim 10^{10}$ & $\Phi_{f(1)}^{max} \sim 10^8$  \\ \hline
\multirow{2}{*}{} 2 & A & $\delta_{i(2)*}'^2 \approx 0.314$ & $\omega'_{i(2)} \approx 102.582$ & $\Phi_{i(2)}^{max} \approx 1.47 \cdot 10^9$ & $\Phi_{i(2)}^{max} \approx 1.47 \cdot 10^5$ \\ \cline{2-6}
                  & B & $\delta_{f(2)*}'^2 = 14.317$ & $\omega'_{f(2)} \approx 97.553$ & $\Phi_{f(2)}^{max} \approx 5.55 \cdot 10^7$ & $\Phi_{f(2)}^{max} \approx 5.55 \cdot 10^3$  \\ \hline
\multirow{2}{*}{} 3 & A & $\delta_{i(3)*}'^2 = 0.49$ & $\omega'_{i(3)} \approx 107.336$ & $\Phi_{i(3)}^{max} \approx 1.00 \cdot 10^7$ & $\Phi_{i(3)}^{max} \approx 10.0$  \\ \cline{2-6}
                  & B & $\delta_{f(3)*}'^2 = 29.362$ & $\omega'_{f(3)} \approx 103.621$ & $\Phi_{f(3)}^{max} \approx 2.34 \cdot 10^5$ & $\Phi_{f(3)}^{max} \approx 0.234$  \\ \hline
\end{tabular}
\label{Table 1}
\end{table*}

It is important to underline that for the first resonance the frequency and the cross-section of the process decrease with the increase of the parameters $\delta_{i, f}'^2$ (see Table 1). However, for the second and third resonances the functions $F_{i(r)}^{max}$ and $F_{f(r)}^{max}$ realize maximums at the specific $\delta_{i(r)*}'^2$ and $\delta_{f(r)*}'^2$ values. Thus, the calculations derive that starting from the second resonance there are distinctive frequencies for which the electron radiates with maximal probability. Additionally, the specified frequencies are various for the channels A and B and exceed the frequencies of the radiation of the first resonance (see Table 1). The most probable frequencies of the higher resonances dispose outside the emission area of the first resonance, however, notwithstanding the small-scaled grade of the cross-sections of the highest resonances in contrast to the first resonance their appearance is nonetheless substantial (see Table 1).

\section{\label{sec:level1}Conclusion}
The accomplished investigation of the process of the resonant SB of ultrarelativistic electrons in the fields of a nucleus and a weak quasimonochromatic electromagnetic wave summarizes the following fundamental results:\\
1. The resonant SB effect for the channels A and B effectively splits into two subsequential first-order procedures with accordance to the fine structure constant: the process of spontaneous photon radiation by an electron as a result of the absorption of $r$ - photons of a wave (the laser-stimulated Compton-effect) and the process of electron scattering on a nucleus in the field of a wave (the laser-assisted Mott scattering process).\\
2. The research utilizes the characteristic energy of the process $E_r = E_1 / r$ \eqref{eq:40} where $r = 1, 2, 3,...$ - is the number of resonance. The represented value has the $E_1 \sim 10^2 \: GeV$ order of magnitude for the first resonance within the range of the optical frequencies. Moreover, the article scrutinized the high-energy resonant SB of ultrarelativistic electrons with energy $E_i \gtrsim E_1$ when the initial/final electrons and spontaneous photon advance within a narrow angle cone and diverge considerably from the direction of light wave propagation.\\
3. The angle of the spontaneous photon radiation complementary to the momenta of the initial electron (for the channel A) and of the final electron (for the channel B) defines the resonant frequency that essentially depends on the number of the absorbed photons of a wave (the resonance number). Therefore, for the zero emission angle the frequency of a spontaneous photon obtains a maximum value. Consequently, with increase of the radiation angle the resonant frequency decreases. Furthermore, the indicated magnitude substantially depends on the selection of the interaction channel and the number of resonance.\\
4. The distribution of the resonant differential cross-section on the emission angles of a spontaneous photon for the higher resonances $(r = 2, 3,...)$ specifies a distinctive maximum in contrast to the corresponding distribution for the first resonance. Moreover, the most probable frequencies of a spontaneous photon increase with higher resonance numbers.\\
5. The work proposes that for the laser wave intensity of $I \lesssim 10^{16} \div 10^{17} \: W/cm^2$ and the initial electrons energy $E_i = 125.1 \: GeV$ the resonant frequencies of a spontaneous photon for the channels A and B within the conditions of the first, second and third resonances delineate considerable magnitudes of the differential cross-section from $\sim 10^{12}$ for the first resonance of the channel A to the $\sim 10^5$ (in the units of $\alpha Z^2 r_e^2$) for the third resonance of the channel B (see Table 1).\\
In conclusion, numerous scientific facilities with specialization in pulsed laser radiation (SLAC, FAIR, XFEL, ELI, XCELS) may experimentally verify the computational estimations.\\
\\
\\
\\
\\
\\
\\
\\
\\
\\
\\
\\
\\
\\
\\
\\
\\
\\
\\

%\bibliography{listofreferences}

\end{document}